\documentclass{aa}  

\usepackage{graphicx}
\usepackage{txfonts}
\usepackage[colorlinks=true,linkcolor=blue,citecolor=blue]{hyperref}

\usepackage{url} 
\usepackage{xcolor} 
\usepackage{soul} 
\usepackage{siunitx} 

\definecolor{bluehl}{rgb}{0.75,0.75,1}

\begin{document}

   \title{The mass-dependent UVJ diagram at cosmic noon}

   \subtitle{A challenge for galaxy evolution models and dust radiative transfer}

   \author{Andrea Gebek
          \inst{1},
          Benedikt Diemer
          \inst{2},
          Marco Martorano
          \inst{1},
          Arjen van der Wel
          \inst{1},\\
          Lara Pantoni,
          \inst{1, 3}
            Maarten Baes
          \inst{1},
          Austen Gabrielpillai
          \inst{4},
          Anand Utsav Kapoor
          \inst{1},\\
          Calvin Osinga
          \inst{2},
          Angelos Nersesian
          \inst{1, 5},
          Kosei Matsumoto
          \inst{1},
          and Karl Gordon
          \inst{1,6}
          }

   \institute{Sterrenkundig Observatorium, Universiteit Gent,
              Krijgslaan 281 S9, 9000 Gent, Belgium\\
              \email{andrea.gebek@ugent.be}
         \and
             Department of Astronomy, University of Maryland,
             College Park, MD 20742, USA
        \and
            Institute of Astronomy, KU Leuven, Celestijnenlaan 200D,
            3001 Leuven, Belgium
         \and
            Department of Astrophysics, The Graduate Center, City University of New York, 365 5th Ave, New York, NY 10016, USA
         \and
            STAR Institute, Quartier Agora - Allée du six Août,
            19c B-4000 Liège, Belgium
        \and
            Space Telescope Science Institute, 3700 San Martin Drive,
            Baltimore, MD 21218, USA
             }

   \date{Received XXX; accepted YYY}

 
  \abstract
   {The UVJ color-color diagram is a widely used diagnostic to separate star-forming and quiescent galaxies. Observational data from photometric surveys reveal a strong stellar mass trend, with higher-mass star-forming galaxies being systematically more dust-reddened.}
   {We analyze the UVJ diagram in the TNG100 cosmological simulation at cosmic noon ($z\approx2$). Specifically, we focus on the trend between UVJ colors and mass, which has not been reproduced in any cosmological simulation thus far.}
   {We applied the SKIRT dust radiative transfer code to the TNG100 simulation to generate rest-frame UVJ fluxes. These UVJ colors were then compared to observational data from several well-studied extragalactic fields from the CANDELS/3D-HST programs, augmented by recent JWST/NIRCam photometry.}
   {Quiescent and low-mass ($M_\star\lesssim10^{10.5}\,\mathrm{M}_\odot$) galaxies at cosmic noon do not require significant levels of dust reddening, as opposed to massive ($M_\star\gtrsim10^{11}\,\mathrm{M}_\odot$) star-forming galaxies. An extensive range of possible dust models fall short of the required dust reddening in $\textit{V}-\textit{J}$ color for massive star-forming galaxies, with the simulated galaxies being too blue by $\approx0.9\,\mathrm{mag}$.}
   {We find that only variations in the star-to-dust geometries of the simulated galaxies are able to yield $\textit{V}-\textit{J}$ colors that are red enough to match the observations. A toy model with isolated dust screens around younger stellar populations (with ages below $\sim1\,\mathrm{Gyr}$) can reproduce the observational data, while all "conventional" dust radiative transfer models (where the dust distribution follows the metals in the interstellar medium) fail to achieve the required $\textit{V}-\textit{J}$ colors.}

    \titlerunning{The mass-dependent UVJ diagram}
    \authorrunning{Andrea Gebek et al.}
   \maketitle



\section{Introduction}

The spectral energy distributions (SEDs) of galaxies are the outcome of the cumulative emission from their stellar populations, reprocessed by dust and gas in their interstellar medium (ISM). Through SED fitting (\citealt{Walcher2011}; \citealt{Conroy2013}; \citealt{Pacifici2023}), it is possible to infer various stellar and ISM properties and to inform galaxy evolution theories. One particularly important element of these theories is the star-formation main sequence (see \citealt{Popesso2023} and references therein). Galaxies evolve along this main sequence by actively forming stars, thus, they are able to maintain a younger, blue stellar population. If galaxies shut down their star formation and become quiescent (e.g., \citealt{Peng2010}), their stellar populations become older and end up transitioning to redder colors. These distinct evolutionary pathways are hence reflected in the optical colors of galaxies. Indeed, low-redshift surveys have observed such an optical color bimodality (\citealt{Strateva2001}; \citealt{Baldry2004}; \citealt{Taylor2015}), which has been identified as a critical requirement for galaxy evolution models (\citealt{Cole2000}; \citealt{Nelson2018}).

However, the optical colors of galaxies are not only affected by their stellar populations, but also by dust attenuation. Dust grains in the ISM absorb and scatter starlight more effectively at shorter wavelengths, so the dust makes galaxies appear redder. While this dust reddening is not strong enough to obfuscate the distinctive colors of the star-forming and quiescent galaxy populations in the Local Universe, the red galaxy population is increasingly polluted with dusty star-forming galaxies (DSFGs) at higher redshifts (\citealt{Whitaker2011}). To break this degeneracy between quiescent and dusty star-forming galaxies, a second color needs to be used (e.g., \citealt{Labbe2005}). A particularly effective combination to separate star-forming and quiescent galaxies consists of the $\textit{V}\,(0.55\,\mu\mathrm{m})-\textit{J}\,(1.24\,\mu\mathrm{m})$ and $\textit{U}\,(0.35\,\mu\mathrm{m})-\textit{V}\,(0.55\,\mu\mathrm{m})$ colors, and this color-color plot (the so-called UVJ diagram) is commonly used in galaxy surveys for $z\lesssim3$ to separate the two galaxy populations (\citealt{Wuyts2007}; \citealt{Williams2009}; \citealt{Whitaker2011}; \citealt{Muzzin2013}; \citealt{Fumagalli2014}; \citealt{Fang2018}). In this UVJ diagram, quiescent galaxies populate a distinct region with red $\textit{U}-\textit{V}$ and blue $\textit{V}-\textit{J}$ colors (top-left corner in the UVJ diagram), while dust-obscured star-forming galaxies are red in both $\textit{U}-\textit{V}$ and $\textit{V}-\textit{J}$ (top right corner).

The UVJ diagram is also a useful diagnostic in the context of cosmological, hydrodynamical simulations to assess their realism (\citealt{Trayford2017}; \citealt{Dave2017}; \citealt{Donnari2019}; \citealt{Baes2024}; \citealt{Gebek2024}) and to study how the observed UVJ diagram emerges (\citealt{Akins2022}). Studying the UVJ diagram for simulated galaxies requires a procedure to compute fluxes from the simulated stellar and ISM properties. This procedure ranges from stellar population synthesis methods combined with dust attenuation laws (\citealt{Zuckerman2021}; \citealt{Nagaraj2022}) to more sophisticated dust attenuation modeling, depending on the line-of-sight gas column density (\citealt{Dave2017}; \citealt{Nelson2018}; \citealt{Donnari2019}) and three-dimensional (3D) dust radiative transfer (\citealt{Trayford2017}; \citealt{Akins2022}; \citealt{Baes2024}; \citealt{Gebek2024}). In the Local Universe, \citet{Gebek2024} showed that the UVJ diagram of simulated galaxies from the TNG100 simulation broadly agrees with observational data from the Galaxy And Mass Assembly (GAMA) survey (\citealp{Driver2009, Driver2011}; \citealt{Liske2015}; \citealt{Baldry2018}; \citealt{Driver2022}).

However, at cosmic noon ($z=2$), the population of increasingly dust-reddened massive star-forming galaxies could so far not be reproduced with dust radiative transfer applied to cosmological simulations. \citet{Donnari2019} find that the TNG100 UVJ diagram roughly fits observational data from the 3D-HST survey (\citealt{Brammer2012}; \citealt{Skelton2014}; \citealt{Momcheva2016}), but is missing the observed population of very red star-forming galaxies. \citet{Akins2022} showed that using the SIMBA cosmological simulation (\citealt{Dave2019}), with its on-the-fly dust model leads to some simulated galaxies being very dust-reddened, thereby populating the top right corner of the UVJ diagram, but these are mostly low-mass galaxies ($M_\star\lesssim10^{10.5}\,\mathrm{M}_\odot$). On the other hand, in observational studies only high-mass galaxies ($M_\star\gtrsim10^{10.5}\,\mathrm{M}_\odot$) populate this region. In summary, the strong stellar mass trend in the observed UVJ colors at cosmic noon (\citealt{Skelton2014}) has not yet been reproduced in the framework of galaxy evolution models.

In this work, we carry out an in-depth analysis of the UVJ diagram as a function of stellar mass in observations and cosmological simulations to perform a more stringent test on galaxy evolution models coupled with dust radiative transfer. Because low-mass galaxies dominate the sample in number, it is crucial to split the galaxy population in different mass bins, as the strong trend of UVJ colors with stellar mass is washed out when showing the UVJ diagram for the full galaxy population. Moreover, we can use this mass-resolved UVJ diagram to gain insight into the galaxy population and, especially, the dust properties of observed galaxies at cosmic noon. To this end, we use observational data based on the CANDELS/3D-HST programs supplemented with JWST/NIRCam photometry at $1.8\leq z\leq2.2$. We then compare these observations in terms of the mass-resolved UVJ-diagram to simulated galaxies from TNG100, postprocessed with 3D dust radiative transfer using SKIRT (\citealt{Baes2011}; \citealp{Camps2015, Camps2020}). 

The outline of this paper is as follows: We introduce the observational and TNG100 datasets and the methods to compute rest-frame UVJ fluxes in Sect.~\ref{sec:Methods}. The mass-resolved UVJ diagram is discussed in Sect.~\ref{sec:MassResolvedUVJ} and we analyze the massive dusty star-forming galaxy population whose red $\textit{V}-\textit{J}$ colors are particularly challenging to reproduce in cosmological simulations in Sect.~\ref{sec:DSFGs}. We discuss our results in the context of other cosmological simulations and recent observational insights in Sect.~\ref{sec:Discussion} and present our conclusions in Sect.~\ref{sec:Conclusion}. We have adopted a flat $\Lambda$CDM cosmology, with parameters measured by the Planck satellite (\citealt{Planck2016}), consistent with the IllustrisTNG cosmology. We use the AB magnitude system (\citealt{Oke1983}) throughout this study.

\section{Methods}\label{sec:Methods}

To analyze the mass-resolved UVJ diagram, the stellar masses and rest-frame UVJ fluxes are required for both the observed and simulated galaxies. For the observational data, we rely on a recently published catalog from the DAWN JWST Archive (DJA\footnote{\url{https://dawn-cph.github.io/dja/index.html}}) described in Sect.~\ref{sec:obsData}. Furthermore, we have adopted TNG100 as our simulation dataset (see Sect.~\ref{sec:Discussion} for a comparison to other cosmological simulations). We describe the IllustrisTNG simulations and our method to compute dust-attenuated fluxes in a post-processing step with 3D dust radiative transfer in Sect.~\ref{sec:TNG100}.

\subsection{Observational data: JWST/NIRCam}\label{sec:obsData}

Our observational dataset consists of a publicly available galaxy catalog\footnote{\url{https://s3.amazonaws.com/aurelien-sepp/full-good_morpho-phot.fits.gz}} from the DJA. This catalog is based on \textit{James Webb} Space Telescope (JWST) NIRCam imaging mosaics (\citealt{Valentino2023}) in the PRIMER-COSMOS (version 7.0), PRIMER-UDS (v7.2), GOODS-S (v7.2), GOODS-N (v7.3), and CEERS (v7.2) fields. These survey regions are designed to overlap with those observed by the \textit{Hubble} Space Telescope (HST) during the Cosmic Assembly Near-infrared Deep Extragalactic Legacy Survey (CANDELS; \citealt{Koekemoer2011}) and 3D-HST (\citealt{Brammer2012}; \citealt{Skelton2014}; \citealt{Momcheva2016}) programs. We briefly describe the salient features of this catalog in the following.

The DJA catalog sources are detected on noise-equalized images of the three longest-wavelength wide NIRCam filters (F277W, F356W, and F444W). The photometry was extracted within circular apertures with diameters of \SI{0.5}{\arcsecond} (corresponding to 4.3\,kpc at $z=2$) and then corrected to total fluxes based on the flux within an elliptical Kron aperture (\citealt{Kron1980}) in the detection image. This means that if radial color gradients outside of a radius of $\approx2\,\mathrm{kpc}$ are present in the galaxies, the aperture-corrected fluxes reported in the catalog are not representative of the true total fluxes. As we show in Appendix~\ref{sec:Observation variations}, using a different observational dataset based only on CANDELS/3D-HST (i.e., without JWST data) with a different aperture correction (in slightly larger apertures with diameters of \SI{0.7}{\arcsecond}) does not affect our conclusions. However, to unambiguously assess how reliably the aperture colors match the total colors, an analysis of the spatially resolved HST/JWST images would be required. This   is beyond the scope of this work.

The DJA catalog\ contains aperture-corrected fluxes in up to 24 bands from JWST/MIRI, JWST/NIRCam, HST/WFC3, and HST/ACS in the wavelength range $0.43-20.8\,\mu\mathrm{m}$. The photometric redshift fitting code EAZY (\citealt{Brammer2008}) is then run for $\approx340000$ objects, measuring redshifts, rest-frame fluxes, and stellar population parameters (e.g., stellar masses, V-band attenuations), which are also included in the catalog. We note that our results hinge in particular on reliable photometric redshifts, because biased redshifts will also affect rest-frame fluxes and stellar masses. We select all objects from the catalog with $M_\star\geq10^{9.5}\,\mathrm{M}_\odot$ and $1.8\leq z\leq2.2$, leading to a sample of 3609 galaxies. In the remainder of this paper, we refer to this observational sample as `JWST/NIRCam' data.

\subsection{Simulation data: TNG100}\label{sec:TNG100}

\subsubsection{The IllustrisTNG simulations}

The IllustrisTNG project (\citealt{Springel2018}; \citealt{Pillepich2018a}; \citealt{Nelson2018}; \citealt{Naiman2018}; \citealt{Marinacci2018}) is a suite of cosmological, magnetohydrodynamical simulations that have been run from $z=127$ to $z=0$. The suite consists of three different volumes with box sizes of approximately 50, 100, and 300 comoving Mpc and is publicly available at \url{www.tng-project.org} (\citealt{Nelson2019}). All simulations were run using the moving-mesh code AREPO (\citealt{Springel2010}) and realized in three to four different resolutions with cosmological parameters set to the 2015 Planck results (\citealt{Planck2016}). For this study, we only made use of the TNG100-1 (hereafter `TNG100') run, which is the highest-resolution version of the $75/h$-cMpc box, with a baryonic mass resolution of $1.4\times10^6\,\mathrm{M}_\odot$. It comprises a suitable compromise between volume and resolution for this study. This is the fiducial simulation of the IllustrisTNG suite since the subgrid parameters (mostly related to galactic winds and black holes processes that are not resolved by the simulation) were chosen at this resolution (\citealt{Pillepich2018b}). In the following, we briefly describe the aspects of the IllustrisTNG model (\citealt{Weinberger2017}; \citealt{Pillepich2018b}) that are most relevant to this study.

Gas radiative processes (primordial and metal-line cooling) allow the gas to cool down to $T\!\sim\!10^4\,\mathrm{K}$. The unresolved cold, dense ISM phase is simulated according to the two-phase model of \citet{Springel2003}, including stochastic star formation for gas with $n_\mathrm{H}>0.106\,\mathrm{cm}^{-3}$. Stellar populations inherit the metallicity of their natal gas cell and follow a Chabrier initial mass function (\citealt{Chabrier2003}). These star particles subsequently affect the surrounding ISM via metal enrichment as well as feedback from supernovae explosions. The IllustrisTNG model furthermore incorporates the formation and merging of supermassive black holes, feedback from active galactic nuclei in a thermal, and a kinetic mode, and the seeding and evolution of magnetic fields.

Subhalos (i.e., galaxies) are identified as gravitationally bound substructures using the SUBFIND algorithm (\citealt{Springel2001}). All galaxy and particle properties are stored at discrete timesteps (snapshots). For this work, we only used snapshot 33, corresponding to $z=2$. We selected all subhalos in this snapshot with total stellar masses of $M_\star\geq10^{9.5}\,\mathrm{M}_\odot$ and removed spurious subhalos that have a "SubhaloFlag" of 1, giving a sample of 6442 galaxies. 

\subsubsection{Dust radiative transfer with SKIRT}

While galaxy properties like stellar masses and star-formation rates are directly available from the simulation data, the broadband fluxes need to be computed in a post-processing step, taking dust attenuation into account. For this task we used the 3D Monte Carlo dust radiative transfer code SKIRT (\citealt{Baes2011}; \citealp{Camps2015, Camps2020}). The methodology to generate broadband fluxes with SKIRT largely follows the strategy adopted by \citet{Trcka2022} and \citet{Gebek2024}, which are, in turn, based on \cite{Camps2016, Camps2018} and \citet{Kapoor2021}. 

In short, SKIRT simulates the emission from star particles using a set of SED templates chosen by the user. Photon packets are then propagated through the dusty ISM, where they get absorbed and scattered. Since dust is not explicitly followed in IllustrisTNG, the dust distribution as well as a dust model that describes the optical properties of the grains also have to be chosen by the user. Lastly, the photon packets are recorded by synthetic instruments to measure the broadband fluxes in the UVJ bands. In the following, we briefly describe the fiducial settings that we chose for our SKIRT simulations. 

For the stellar emission, we followed \citealt{Vijayan2022} and adopted the BPASS version 2.2.1 SED templates (\citealt{Eldridge2017}; \citealt{Stanway2018}) with a Chabrier initial mass function\footnote{We did not expect this choice to affect any of our results significantly. \citet{Akins2022} found a negligible difference of $\sim\!0.03\,\mathrm{mag}$ on the UVJ colors of galaxies from the SIMBA simulation when varying the initial mass function between \citet{Salpeter1955}, \citet{Kroupa2002}, and \citet{Chabrier2003} in their post-processing method.} and an upper limit of $300\,\mathrm{M}_\odot$. These SED templates are parametrized on initial mass, stellar metallicity, and stellar age, which can all be directly obtained from the simulation data. Using the BPASS template library, we can calculate dust-free fluxes for the TNG100 galaxies.

Since young stellar populations are typically enshrouded in their dusty birth clouds (which happens on spatial scales unresolved by large-volume cosmological simulations), post-processing studies of simulated galaxies usually invoke a separate SED template library for young stellar populations ($t\lesssim10\,\mathrm{Myr}$), which effectively corresponds to a subgrid model for the dust and gas in star-forming regions (\citealt{Jonsson2010}). We modeled star-forming regions using the recently developed TODDLERS library (\citealt{Kapoor2023}) for all star particles with ages below 30\,Myr. The TODDLERS library is parametrized on initial mass, stellar metallicity, stellar age, star-formation efficiency, and cloud density. Following \citet{Kapoor2024}, we used the high-PAH-fraction version of TODDLERS with a fixed star-formation efficiency of $2.5\,\%$ and cloud density of $320\,\mathrm{cm}^{-3}$. 

In a SKIRT calculation, each photon packet is launched from a source (star particles in our case) and then propagated through a transfer medium. In our case, this transfer medium corresponds to the resolved dusty ISM which attenuates starlight. Since the modeling of dust grains is not  covered in IllustrisTNG, we followed the canonical assumption and assign dust to gas cells with a constant dust-to-metal mass ratio\footnote{Recent hydrodynamical simulations that incorporate dust grain physics show that the cold ISM exhibits elevated dust-to-metal ratios and larger grains (see Fig. 9 of \citealt{Dubois2024}). Hence, our usage of a fixed $f_\mathrm{dust}$ is an overly simplistic description of the dusty ISM, but we note that it is hard to meaningfully improve upon this assumption without resolving the cold ($T\lesssim10^3\,\mathrm{K}$) ISM phase. We expect that our $f_\mathrm{dust}$ variations discussed in Sect.~\ref{sec:fDust} bracket a more realistic dust distribution in the diffuse ISM.} (but see \citealt{Popping2022} who use a metallicity-dependent $f_\mathrm{dust}$ based on observational results from \citealt{Remy-Ruyer2014}). This dust-to-metal ratio $f_\mathrm{dust}$ varies between studies and is often calibrated to observational data (e.g., \citealt{Camps2016}; \citealt{Vogelsberger2020}; \citealt{Kapoor2021}; \citealt{Trcka2022}). While $f_\mathrm{dust}$ can be constrained in the local Universe (\citealt{deVis2019}; \citealt{Galliano2021}; \citealt{Zabel2021}), at higher redshifts, it becomes increasingly difficult to measure the dust-to-metal ratios due to uncertainties in the metallicity and dust-to-gas ratio measurements. We adopted $f_\mathrm{dust}=0.5$ as our fiducial dust-to-metal ratio and discuss how $f_\mathrm{dust}$ affects the UVJ diagram in Sect.~\ref{sec:fDust}. Since dust grains are typically destroyed in the hot circumgalactic medium of galaxies, we followed \citet{Trcka2022} and only assigned dust to relatively cold and dense gas cells following the criterion from \cite{Torrey2012, Torrey2019}. For the optical properties of the dust grains, we used the THEMIS dust model for the diffuse ISM (\citealt{Jones2017}).

\begin{figure}
    \centering
    \includegraphics[width=\columnwidth]{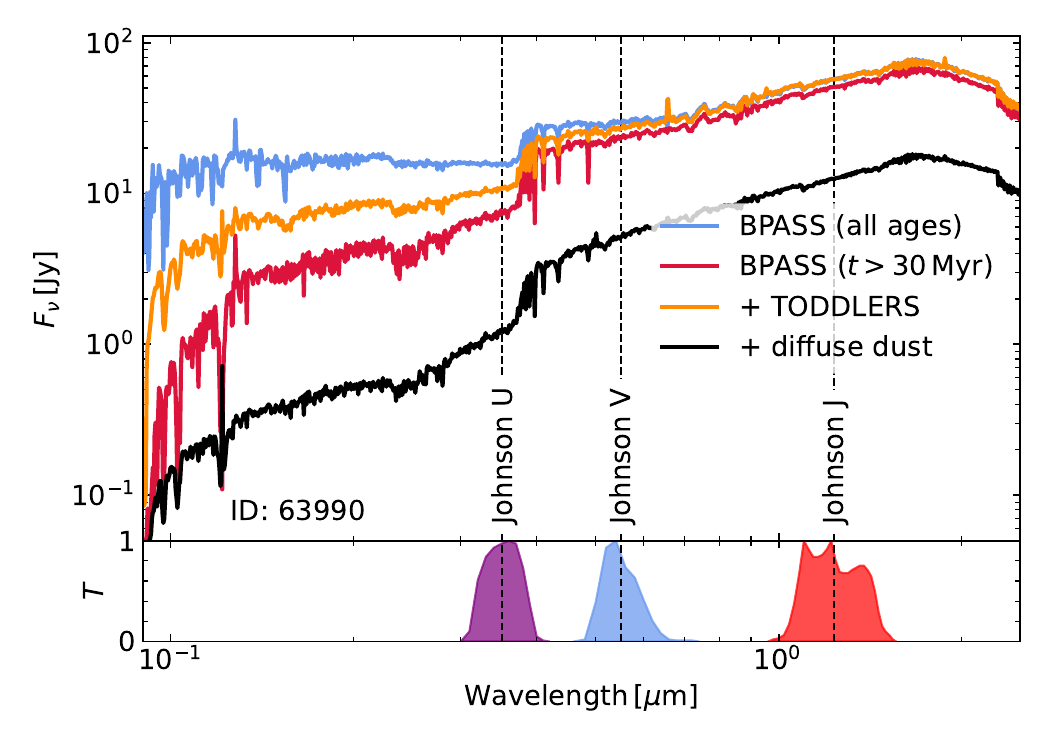}
    \caption{Rest-frame SED for an example TNG100 galaxy at $z=2$ (subhalo ID: 63990, $M_\star\approx10^{11.2}\,\mathrm{M}_\odot$, $\mathrm{SFR}\approx140\,\mathrm{M}_\odot\mathrm{yr}^{-1}$), recorded at a distance of $10\,\mathrm{Mpc}$. The blue line indicates the dust-free SED, i.e., using BPASS for all stellar populations and neglecting the diffuse dust in the ISM. Using only the evolved stellar populations (with ages above $30\,\mathrm{Myr}$) leads to the red SED. The addition of younger stellar populations with TODDLERS templates (which incorporate unresolved dust) gives rise to the orange spectrum. Adding diffuse dust to this orange SED then leads to the final dust-attenuated SED shown in black. The lower panel indicates the (arbitrarily normalized) transmission curves for the Johnson UVJ broadband filters.}
    \label{fig:exampleSED}
\end{figure}

\begin{figure*}
    \centering
    \includegraphics[width=\textwidth]{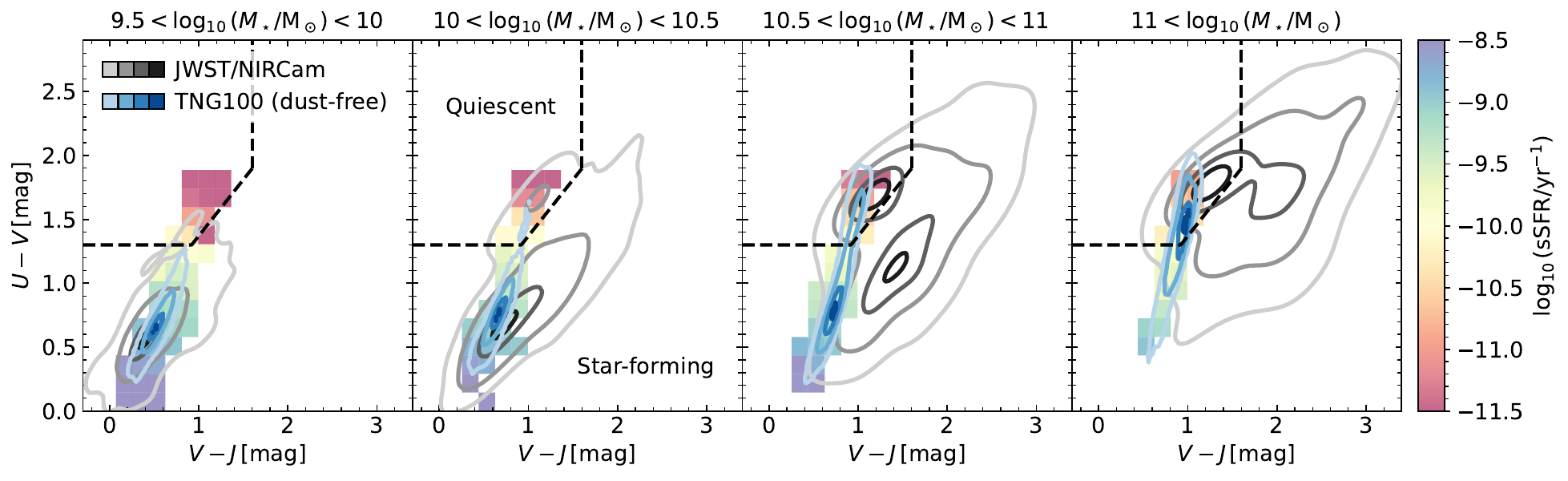}
    \caption{UVJ diagram in bins of stellar mass, for dust-free TNG100 fluxes (blue contours) and observational JWST/NIRCam data (with $1.8\leq z\leq2.2$, grey contours), which are dust-attenuated. The dashed line indicates the demarcation between quiescent and star-forming galaxies from \citealt{Williams2009} in their highest redshift bin ($1\leq z\leq2$). Here, and in all other figures, the contours are estimated from a two-dimensional (2D) kernel density estimate and enclose the densest 5, 25, 70, and 95 percent of the data. The TNG100 distributions are color-coded by their mean logarithmic specific star-formation rates in bins of $\textit{V}-\textit{J}$ and $\textit{U}-\textit{V}$ color. A strong correlation of sSFR with the dust-free UVJ colors of the simulated galaxies is evident. Dust attenuation must shift the simulated massive star-forming galaxies to redder colors in order to reproduce the observational distribution in the UVJ diagram.}
    \label{fig:UVJ_sSFR}
\end{figure*}

Lastly, we describe the most important technical settings that we chose for our SKIRT simulations. Since we did not consider dust emission (our tests indicate that dust emits only at wavelengths above $\approx2.5\,\mu\mathrm{m}$; see also Fig. 1 of \citealt{Gebek2024}), we run SKIRT in its "ExtinctionOnly" mode. We used $10^7$ photon packets for each simulation which are emitted over a wavelength range of $0.29-1.6\,\mu\mathrm{m}$ to cover the rest-frame UVJ bands. Each SKIRT simulation is defined on a specific spatial domain, we used a cube of side length $\max(60\,\mathrm{kpc},10\times R_{\mathrm{half},\star}),$ where $R_{\mathrm{half},\star}$ denotes the stellar half-mass radius of the galaxy. This spatial domain is discretized with an octree grid (\citealt{Saftly2013,Saftly2014}) with a minimum (maximum) refinement level of 5 (10), with the refinement criterion chosen such that cells will be refined until the dust mass in each cell is below $2\times10^{-6}$ times the total dust mass in the galaxy. For a spatial domain with a side length of $60\,\mathrm{kpc}$ (which is the box size for 98.6\,\% of the TNG100 galaxies in our sample), the maximum refinement level corresponds to a spatial resolution of $59\,\mathrm{pc}$. This should be sufficient given that the minimum value of the gravitational softening length for gas cells in TNG100 is $185\,\mathrm{pc}$. The photon packets are registered in a synthetic instrument\footnote{The orientation of the instrument is set to observe galaxies along the $z$-axis of the TNG100 box. This leads to random viewing angles between the galaxy and the observer as in the observational data.} to measure the fluxes in the UVJ bands in a circular aperture with a radius of $30\,\mathrm{kpc}$. We  tested apertures of $5\times R_{\mathrm{half},\star}$ for the galaxies where $R_{\mathrm{half},\star}>6\,\mathrm{kpc}$, but we found the differences in the UVJ colors to be negligible. 

We show an example SED generated with SKIRT in Fig.~\ref{fig:exampleSED} for a massive ($M_\star\approx10^{11.2}\,\mathrm{M}_\odot$), star-forming ($\mathrm{SFR}\approx140\,\mathrm{M}_\odot\mathrm{yr}^{-1}$) TNG100 galaxy at $z=2$. The SED is shown in the galaxy rest-frame from $0.09-2.5\,\mu\mathrm{m}$, up to which dust emission can safely be ignored. The dust-free SED generated by using BPASS for all stellar populations and neglecting diffuse dust is shown in blue. The flux due to the evolved stellar populations alone is shown in red, this corresponds to the composite SED of all stellar populations with ages above $30\,\mathrm{Myr}$ modeled with BPASS. Adding the younger stellar populations (modeled with TODDLERS) mostly adds to the flux bluewards of the 4000\,\AA-break as shown with the orange SED. The difference between the orange and the blue SED is due to unresolved dust from the dusty birth clouds incorporated in TODDLERS. Lastly, the resolved dust attenuates the orange SED and leads to the final spectrum shown in black. The attenuation is stronger at shorter wavelengths, but we note that there is significant absorption, even at $2.5\,\mu\mathrm{m}$. The Johnson UVJ filters used throughout this work to compute broadband fluxes are indicated in the lower panel of Fig.~\ref{fig:exampleSED}.

Figure~\ref{fig:exampleSED} illustrates the various galaxy components that determine the global UVJ fluxes. The U-band flux has a significant contribution from young stellar populations and is strongly attenuated, while the V and J bands are dominated by the evolved stellar populations (and significantly attenuated by dust). This means that the $\textit{U}-\textit{V}$ and $\textit{V}-\textit{J}$ colors arise due to a convolution of the intrinsic (dust-free) colors of the composite stellar population, modulated by dust attenuation which reddens these colors depending on the properties of the dust grains and the star-to-dust geometry (\citealt{Narayanan2018}; \citealt{Salim2020}).

With the SKIRT post-processing, we can obtain rest-frame dust-free (using BPASS for all star particles) and dust-attenuated fluxes (using the BPASS and TODDLERS templates and incorporating resolved dust) in the UVJ bands for the 6442 galaxies in our TNG100 sample. We caution that many of our SKIRT post-processing settings are ambiguous choices and reflect uncertainties for instance in present-day stellar evolution models, stellar libraries, or dust grain properties. Hence, these SKIRT settings constitute a source of systematic uncertainty. We  tested many alterations of our fiducial SKIRT post-processing method  (shown in Appendix~\ref{sec:SKIRT variations}). We find that variations in the dust treatment only marginally affect the UVJ colors, while the choice of SED templates affects the UVJ colors up to $\lesssim0.4\,\mathrm{mag}$. Importantly, our conclusions are robust to all SKIRT post-processing variations that we explored.

\section{The mass-dependent UVJ diagram}\label{sec:MassResolvedUVJ}

\begin{figure*}
    \centering
    \includegraphics[width=\textwidth]{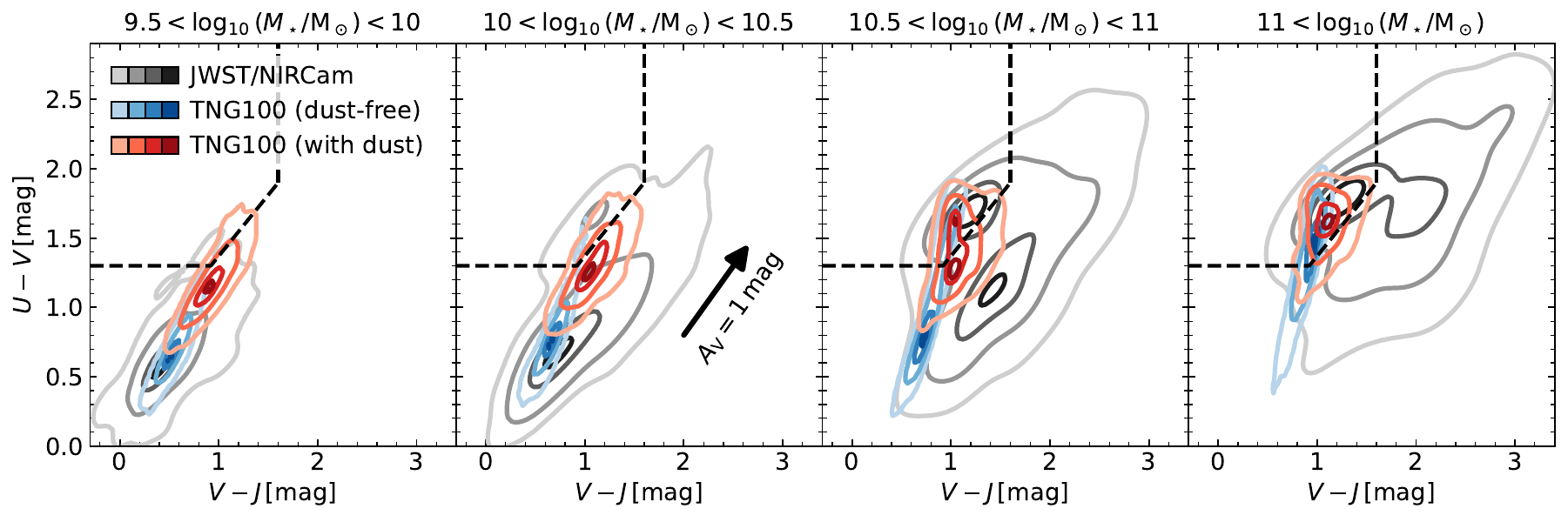}
    \caption{Impact of dust attenuation on the UVJ colors of TNG100 galaxies. The blue contours show dust-free fluxes, while the red contours incorporate dust attenuation from unresolved dust (using the TODDLERS templates for star-forming regions) and resolved dust (adopting a dust-to-metal ratio of 50\,\% in the ISM). The arrow in the first panel indicates the reddening in $\textit{V}-\textit{J}$ and $\textit{U}-\textit{V}$ according to the THEMIS dust model with an extinction in the V-band of one magnitude. Dust reddening is overly effective at low stellar masses while for massive star-forming galaxies the reddening in $\textit{V}-\textit{J}$ is insufficient to reproduce the observations.}
    \label{fig:UVJ_diffuseDust}
\end{figure*}

Equipped with the stellar masses and UVJ rest-frame colors for the observational data and TNG100, we can then analyze the UVJ diagram in bins of stellar mass. The trends in the UVJ diagram are governed by the properties of the stellar populations, modulated by dust attenuation depending on the amount of dust, the optical properties of the dust grains, and the star-to-dust geometry. To isolate the various effects at play, we first analyze the colors and star-formation rates of simulated galaxies without dust in Sect.~\ref{sec:Dust-free SFR}. We then discuss the impact of dust attenuation on the UVJ diagram in Sect.~\ref{sec:fDust}.

\subsection{Dust-free UVJ diagram}\label{sec:Dust-free SFR}

We show the dust-free UVJ color distributions for the TNG100 galaxies in Fig.~\ref{fig:UVJ_sSFR} (blue contours). The galaxies form a relatively narrow and steep sequence and the fraction of red galaxies (in $\textit{U}-\textit{V}$) increases with stellar mass. The observational data are indicated by the grey contours, but we do not expect this distribution to match the simulation results here since the JWST/NIRCam colors are dust-attenuated. It would be possible to retrieve "dust-free" colors for the observed galaxies with SED fitting to compare them to the intrinsic colors of the simulated galaxies. Indeed, \citet{Nagaraj2022} find a good agreement between intrinsic TNG100 and dust-free 3D-HST colors (their Fig. 3). However, we avoided this dust-free comparison because the SED fitting method incorporates many assumptions (e.g., simplified star-to-dust geometries), which could potentially introduce systematic biases on the derived dust-free colors.

To analyze how the different intrinsic UVJ colors emerge for the simulated galaxies, we also show the average logarithmic specific star-formation rate (sSFR) in bins of $\textit{V}-\textit{J}$ and $\textit{U}-\textit{V}$ colors. The $\textit{U}-\textit{V}$ color is strongly anticorrelated with sSFR such that galaxies with $\mathrm{sSFR}\lesssim10^{-10}\,\mathrm{yr}^{-1}$ fall into the quiescent area of the UVJ diagram, above the UVJ demarcation line of \citet[see also Fig. 6 of \citealt{Baes2024}]{Williams2009}. Remarkably, the quiescent galaxies in TNG100 are already located at the locus of the JWST/NIRCam galaxies in the quiescent region, even without incorporating any dust into the simulated galaxies. This finding agrees with observational studies of nearby galaxies (\citealt{DeVis2017}) and distant galaxies up to $z\sim5$ (\citealt{Rowlands2014}; \citealp{Donevski2020,Donevski2023}) which find a strong positive correlation between sSFR and dust-to-stellar mass ratio (specific dust mass), although the relation flattens out at $\mathrm{sSFR}\gtrsim10^{-10}\,\mathrm{yr}^{-1}$ for nearby galaxies (\citealt{Remy-Ruyer2014}; \citealt{Galliano2021}). Hence, we do not expect significant dust reddening for the JWST/NIRCam galaxies in the quiescent region.

The quiescent\footnote{We adopted an sSFR cut of $10^{-10.6}\,\mathrm{yr}^{-1}$; see Sect.~\ref{sec:DSFGs} for the justification of this threshold.} TNG100 galaxies in our sample have mass-weighted ages of $1.35^{+0.30}_{-0.27}\,\mathrm{Gyr}$ and mass-weighted metallicities of $0.022^{+0.004}_{-0.003}$ (quoting the medians and 16-84 percentiles). Upon our examination of the colors of the BPASS templates, we find that a simple stellar population (SSP) at this median age and metallicity exhibits a $\textit{V}-\textit{J}$ color of $0.95\,\mathrm{mag}$ and a $\textit{U}-\textit{V}$ color of $1.78\,\mathrm{mag}$. These SSP colors broadly align with the locus of the quiescent JWST/NIRCam galaxies.

\subsection{Dust attenuation in the UVJ diagram}\label{sec:fDust}

We  then considered the effect of dust attenuation on the UVJ colors of the TNG100 galaxies. We incorporate local dust attenuation from star-forming regions by using TODDLERS for young star particles with ages below $30\,\mathrm{Myr}$. Dust attenuation in the diffuse ISM is modeled using a constant dust-to-metal ratio of $f_\mathrm{dust}=50\,\%$. This parameter describes the fraction of metals locked into dust for each gas cell and, hence, it sets the overall amount of resolved dust in the ISM given the gas cell metal masses predicted directly by TNG100. The dust-attenuated UVJ colors are shown in Fig.~\ref{fig:UVJ_diffuseDust} as red contours. For comparison, we also show the dust-free TNG100 colors as blue contours.

As shown in Appendix~\ref{sec:SKIRT variations}, the impact on the UVJ colors of the unresolved dust alone is minor. When neglecting the diffuse dust in the ISM, going from BPASS-only to BPASS-TODDLERS colors, the $\textit{U}-\textit{V}$ ($\textit{V}-\textit{J}$) color is reddened\footnote{When calculating differences in TNG100-SKIRT colors, we always compute the absolute values of the differences in $\textit{V}-\textit{J}$ and $\textit{U}-\textit{V}$ colors for all TNG100 galaxies. We then compute the median of these differences in all stellar mass bins used in Fig.~\ref{fig:UVJ_diffuseDust} separately and quote the largest median value.\label{footnote}} by $\lesssim0.19\,\mathrm{mag}$ ($\lesssim0.04\,\mathrm{mag}$).

On the other hand, the resolved dust in the ISM leads to substantial reddening in both the $\textit{V}-\textit{J}$ and $\textit{U}-\textit{V}$ colors, with the magnitude and direction of the reddening vector depending on the amount of dust, the dust grain properties, and the star-to-dust geometry. In the SKIRT post-processing step, the dust grain properties are fixed according to the THEMIS dust model. The arrow in the first panel of Fig.~\ref{fig:UVJ_diffuseDust} indicates the reddening due to a dust screen with a V-band extinction of one magnitude ($A_\mathrm{V}=1\,\mathrm{mag}$). The direction of this vector will be modulated depending on the star-to-dust geometry in the simulated galaxies (\citealt{Narayanan2018}; \citealt{Akins2022}).

With our fiducial choice of $f_\mathrm{dust}=50\,\%$, the addition of resolved dust yields a dust reddening of $\lesssim0.35\,\mathrm{mag}$ ($\lesssim0.37\,\mathrm{mag}$) in $\textit{V}-\textit{J}$ ($\textit{U}-\textit{V}$). Importantly, the dust reddening due to the diffuse ISM saturates already at low $f_\mathrm{dust}$ values. In $\textit{V}-\textit{J}$ color (the behaviour is similar for $\textit{U}-\textit{V}$), going from $f_\mathrm{dust}=0$ to $f_\mathrm{dust}=0.1$ reddens the galaxies by $\lesssim0.28\,\mathrm{mag}$, from $f_\mathrm{dust}=0.1$ to $f_\mathrm{dust}=0.5$ by $\lesssim0.14\,\mathrm{mag}$, and from $f_\mathrm{dust}=0.5$ to $f_\mathrm{dust}=1$ only by $\lesssim0.05\,\mathrm{mag}$. This well-known effect arises because increased dust content amplifies the contribution of scattered light at bluer wavelengths and the obscured stars also become fainter, thereby contributing less dust-reddened light to the global fluxes (\citealt{Witt1992}).

Comparing the dust-attenuated TNG100 colors to the JWST/NIRCam data in Fig.~\ref{fig:UVJ_diffuseDust}, we note that the quiescent galaxies are hardly dust-reddened, which is best visible in the stellar mass range $10.5<\log_{10}(M_\star/\mathrm{M}_\odot)<11$ and agrees with the observed UVJ colors. On the other hand, for the star-forming galaxies we find some discrepancies between TNG100 and the observational data: At low stellar masses, the TNG100 galaxies are significantly dust-reddened. However, the JWST/NIRCam UVJ diagram suggests negligible dust reddening for these low-mass star-forming galaxies. At the high-mass end, we find that the simulated galaxies are broadly compatible to the observational data in $\textit{U}-\textit{V}$, but are significantly too blue in $\textit{V}-\textit{J}$. This insufficient dust reddening for the massive, star-forming galaxies leads to a contamination\footnote{To quantify the contamination of the quiescent region, we consider the fraction of star-forming galaxies with $M_\star\geq10^{11}\,\mathrm{M}_\odot$ in the quiescent region in our fiducial dust model. We classify all TNG100 galaxies with an sSFR above $10^{-10.6}\,\mathrm{yr}^{-1}$ as star-forming (see Sect.~\ref{sec:DSFGs} for our motivation for this specific sSFR threshold value). Using the UVJ selection of \citet{Williams2009}, we find a contamination with star-forming galaxies in the quiescent region of $76\,\%$ for TNG100.} of the quiescent region at $M_\star\geq10^{10.5}\,\mathrm{M}_\odot$ with star-forming TNG100 galaxies. 

\begin{figure}
    \centering
    \includegraphics[width=\columnwidth]{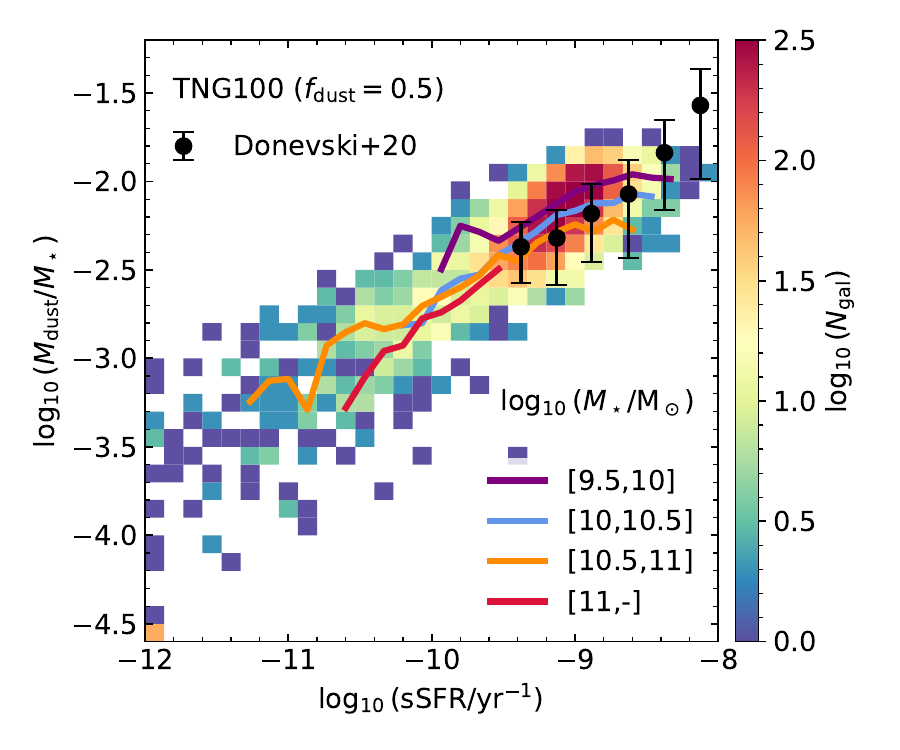}
    \caption{Distribution of TNG100 galaxies as a function of sSFR and specific dust mass, adopting our fiducial dust-to-metal ratio of 0.5. Colored lines indicate running medians in different stellar mass bins. To guide the eye we also show observational data (medians and 16th-84th percentile ranges) from \citet{Donevski2020} for a sample of star-forming galaxies with $0.5<z<5.5$, confirming that the TNG100 dust masses are plausible.}
    \label{fig:Mdust}
\end{figure}

\begin{figure*}
    \centering
    \includegraphics[width=\textwidth]{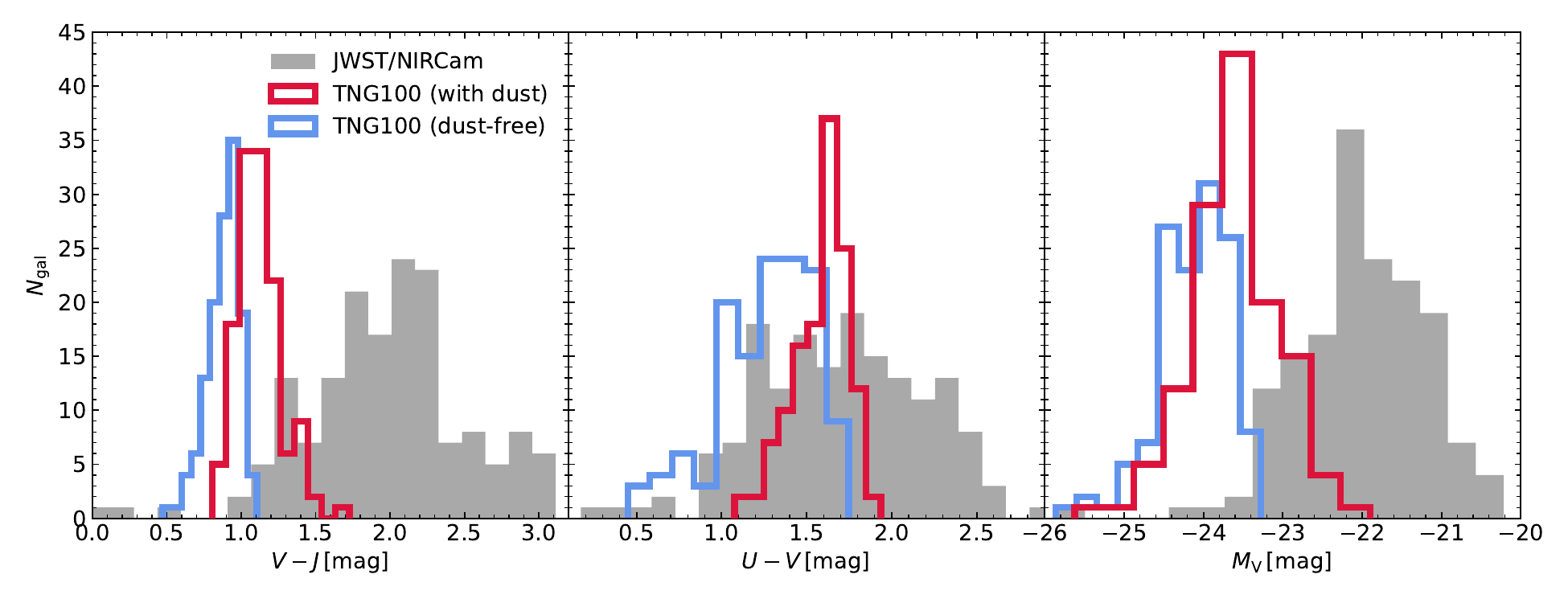}
    \caption{Distributions of massive star-forming galaxies in the JWST/NIRCam and TNG100 datasets. Shown are for all galaxy samples the $\textit{V}-\textit{J}$ and $\textit{U}-\textit{V}$ color distributions (left and center panels, respectively) as well as the absolute V-band magnitudes (right panel). All samples are selected on stellar mass ($M_\star\geq10^{11}\,\mathrm{M}_\odot$). Additionally, we select star-forming galaxies in JWST/NIRCam according to their position in the UVJ diagram (using the cut from \citealt{Williams2009}). For TNG100, we use an sSFR threshold of $10^{-10.6}\,\mathrm{yr}^{-1}$ to select star-forming galaxies. We show the results for the simulated galaxies without dust (blue histograms) and with dust (red histograms). While the $\textit{U}-\textit{V}$ distributions broadly match, we find that the simulated galaxies are too blue in $\textit{V}-\textit{J}$ (by $\approx0.9\,\mathrm{mag}$) and too bright in the V-band (by $\approx1.6\,\mathrm{mag}$), even when incorporating dust attenuation.}
    \label{fig:massiveSFgalaxies_hist}
\end{figure*}

To first order, we can interpret these results by the specific dust masses of the simulated galaxies. We show the specific dust masses of TNG100 with $f_\mathrm{dust}=0.5$ as function of specific star-formation rate in the Fig.~\ref{fig:Mdust}. For TNG100, the dust mass effectively corresponds to the metal mass in the ISM scaled by the $f_\mathrm{dust}$ factor. We also show observational data from \citet{Donevski2020}, based on a study of star-forming galaxies in the redshift range $0.5<z<5.5$, confirming that the dust masses that we used for the radiative transfer post-processing are plausible. Since the quiescent galaxies contain little to no dust, they are hardly dust-reddened in Fig.~\ref{fig:UVJ_diffuseDust}. For the star-forming galaxies, we note that in TNG100 the specific dust mass is highest for the lowest mass bin indicated by the purple line in Fig.~\ref{fig:Mdust}. This leads to significant dust reddening for the low-mass star-forming galaxies, in contrast to the JWST/NIRCam UVJ diagram and observational studies which find these galaxies to exhibit small V-band attenuations (\citealt{Heinis2014}; \citealt{Pannella2015}; \citealt{McLure2018}; \citealt{Zhang2023}). Furthermore, nebular attenuation measurements of cosmic noon galaxies also indicate a positive correlation between stellar mass and dust reddening (\citealt{Shivaei2020}). Hence, with a constant dust-to-metal ratio it is impossible to reproduce the observed UVJ diagram at cosmic noon (see also \citealt{Akins2022} who reach the same conclusion).

While the overly effective dust reddening for low-mass star-forming TNG100 galaxies could easily be remedied by lowering $f_\mathrm{dust}$, increasing the diffuse ISM dust content for massive star-forming galaxies does not lead to redder $\textit{V}-\textit{J}$ colors. Hence, TNG100 fails to reproduce the dust-reddened sequence of massive, star-forming galaxies that is seen in the JWST/NIRCam UVJ diagram. We emphasize that this tension is robust against various systematic uncertainties in the post-processing (see Appendix~\ref{sec:SKIRT variations}), and also persists to other cosmological simulations (SIMBA and EAGLE) as shown in Sects.~\ref{sec:SIMBA} and~\ref{sec:EAGLE}. We analyze this enigmatic galaxy population in Sect.~\ref{sec:DSFGs}, and focus especially on variations in the star-to-dust geometry to understand this discrepancy.

\section{Massive star-forming galaxies at cosmic noon}\label{sec:DSFGs}

We investigate the UVJ colors of massive ($M_\star\geq10^{11}\,\mathrm{M}_\odot$), star-forming galaxies more quantitatively in Fig.~\ref{fig:massiveSFgalaxies_hist} by showing 1D histograms for the $\textit{V}-\textit{J}$ and $\textit{U}-\textit{V}$ colors and for the absolute V-band magnitude ($M_\mathrm{V}$). For the JWST/NIRCam data, the star-forming galaxies are selected by using the UVJ demarcation line from \citet{Williams2009}, leading to a sample of 162 galaxies. For the simulated galaxies, we impose an sSFR cut\footnote{For the TNG100 galaxies, the relation between sSFR and (dust-free) UVJ colors is visualized in Fig.~\ref{fig:UVJ_sSFR}. In the highest-mass bin ($M_\star>10^{11}\,\mathrm{M}_\odot$), the sSFR of TNG100 galaxies with $\textit{U}-\textit{V}=1.2-1.4\,\mathrm{mag}$ (bracketing the demarcation value of $\textit{U}-\textit{V}=1.3\,\mathrm{mag}$) is $\log_{10}(\mathrm{sSFR}/\mathrm{yr}^{-1})=-9.9^{+0.1}_{-0.2}$ (quoting the median and 16-84 percentile values). Since this is a higher sSFR than our imposed cut of $10^{-10.6}\,\mathrm{yr}^{-1}$, there is a sizeable fraction of TNG100 galaxies classified as star-forming that occupy the quiescent region in UVJ space (using dust-free colors).} of $10^{-10.6}\,\mathrm{yr}^{-1}$ leading to a sample of 131 galaxies. This value is determined by subtracting 0.5\,dex from the mode of the TNG100 sSFR distribution in the highest-mass bin (see also Fig.~\ref{fig:mainSequence}). 

The left panel of Fig.~\ref{fig:massiveSFgalaxies_hist} shows one-dimensional (1D) $\textit{V}-\textit{J}$ color histograms for massive, star-forming galaxies. The simulation results are shown without dust (blue histogram) and with dust (red histogram). As noted before, the $\textit{V}-\textit{J}$ colors of the observed galaxies (grey histograms) are much redder than the dust-attenuated TNG100 ones, with the medians of the distributions differing by $\approx0.9\,\mathrm{mag}$. The situation is different for the $\textit{U}-\textit{V}$ color distributions, shown in the center panel of Fig.~\ref{fig:massiveSFgalaxies_hist}. Here, the median dust-attenuated $\textit{U}-\textit{V}$ colors of TNG100 broadly align with JWST/NIRCam. Lastly, we show the distribution of absolute V-band magnitude in the right panel of Fig.~\ref{fig:massiveSFgalaxies_hist}. Notably, the median dust-attenuated TNG100 V-band magnitude is brighter than JWST/NIRCam by $\approx1.6\,\mathrm{mag}$ (a factor of $\approx4.4$ in brightness).

We  verified that these results are robust to the choice of sSFR cut to select star-forming TNG100 galaxies. Variations of the sSFR cut of 0.2\,dex affect the medians of the distributions (shown in Fig.~\ref{fig:massiveSFgalaxies_hist}) by $\lesssim0.05\,\mathrm{mag}$.

\subsection{Dust attenuation estimates}

From Fig.~\ref{fig:massiveSFgalaxies_hist}, it is possible to "reverse-engineer" the average dust attenuation properties of the simulated galaxies that would be required to reproduce the JWST/NIRCam UVJ data. Assuming that the dust-free UVJ fluxes for TNG100 are realistic\footnote{This is an important caveat, which we discuss in more detail in Sect.~\ref{sec:SIMBA} (comparing to dust-free UVJ fluxes from the SIMBA simulation) and in Appendix~\ref{sec:SKIRT variations} (comparing different stellar SED template libraries).} (as also concluded by \citealt{Nagaraj2022}), dust attenuation would need to redden the massive, star-forming galaxy population in TNG100 by $1.11\,\mathrm{mag}$ in $\textit{V}-\textit{J}$, $0.41\,\mathrm{mag}$ in $\textit{U}-\textit{V}$, and extinguish the V-band by $2.05\,\mathrm{mag}$. These values constrain the average dust attenuation curve that is required to match the JWST/NIRCam data. The dust attenuation curve itself emerges from the interplay between the dust extinction curve (determined by the chemical composition of the dust and the dust grain size distribution), the stellar population properties, and the star-to-dust geometry (\citealt{Narayanan2018}; \citealt{Salim2020}). 

\begin{figure}
    \centering
    \includegraphics[width=\columnwidth]{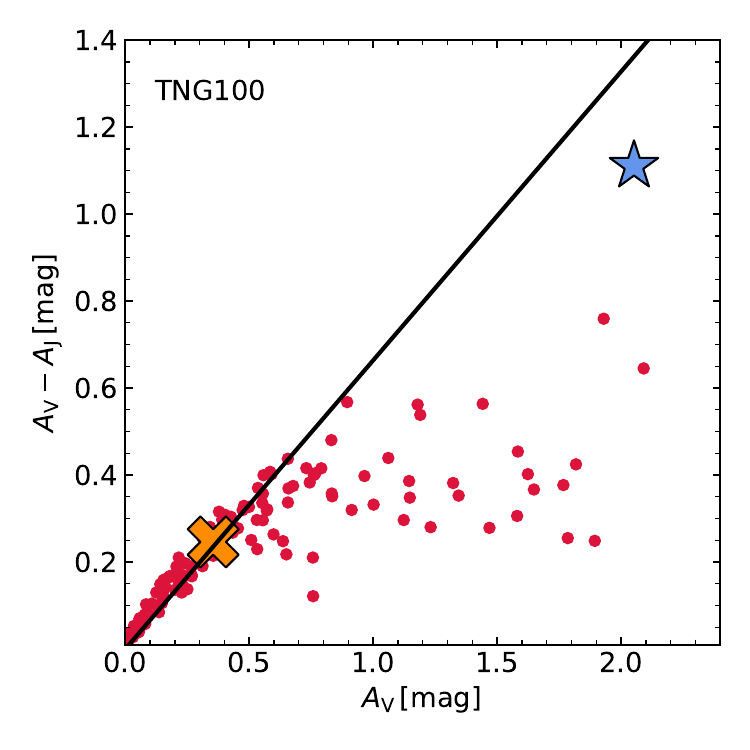}
    \caption{Relationship between galaxy V-band attenuation ($A_\mathrm{V}$) and $\textit{V}-\textit{J}$ reddening ($A_\mathrm{V}\!-\!A_\mathrm{J}$). The straight line shows the expected relation for a thin dust screen with a THEMIS-like extinction curve. The red circles correspond to the attenuation and reddening values in our fiducial dust model for massive, star-forming TNG100 galaxies. The blue star marker indicates the median attenuation and reddening values that the dust-free TNG100 galaxy population would need in order to reproduce the JWST/NIRCam data, while the orange cross marker indicates the actual median values of the TNG100 galaxies. To match the observational data, the attenuation-reddening relation must be almost as steep as in the dust screen model, which is not the case for the TNG100 galaxies postprocessed with dust radiative transfer.}
    \label{fig:AV_AJ_relation}
\end{figure}

For our simulated galaxies, the stellar population properties are fixed and we do not further vary them. The dust extinction curve is set by our assumed dust model in SKIRT. Our fiducial choice is THEMIS, and variations of this dust model (we test the \citealt{Draine2007} and \citealt{Zubko2004} dust models) only marginally affect the UVJ colors of simulated galaxies (see Appendix~\ref{sec:SKIRT variations}). However, all of these dust models are "calibrated" to reproduce data from the Milky Way, and it is not clear how Milky Way dust extinction curves vary compared to star-forming galaxies at cosmic noon. While dust extinction properties measured along nebular sight lines of cosmic noon galaxies resemble Milky Way dust extinction curves (\citealt{Reddy2020}), dust grain properties in the diffuse ISM are still very challenging to constrain (but see \citealt{Shivaei2024}; \citealt{Polletta2024}). Hence, we did not further investigate variations in the dust extinction curve but instead focus on the third driver of the dust attenuation curve: the star-to-dust geometry.

In the simplest star-to-dust geometry where dust exists only in a thin shell around the galaxy (the dust screen model), scattering into the line-of-sight is negligible and the attenuation curve converges to the dust extinction curve. As we held the dust model (THEMIS) and, hence, the extinction curve fixed in our SKIRT post-processing, the shape of the attenuation curve would be the same for all galaxies while the normalization can differ depending on the dust surface density in this dust screen model. This is visualized in Fig.~\ref{fig:AV_AJ_relation}, where we show the dust reddening in $\textit{V}-\textit{J}$ ($A_\mathrm{V}\!-\!A_\mathrm{J}$) as a function of $A_\mathrm{V}$. We focus on these quantities because this is where we encountered substantial tensions between TNG100 and JWST/NIRCam (unlike $\textit{U}-\textit{V}$; see Fig.~\ref{fig:massiveSFgalaxies_hist}).

The dust screen model corresponds to a straight line in Fig.~\ref{fig:AV_AJ_relation}, with a slope of 0.664 dictated by the THEMIS dust extinction curve. The TNG100 results for our fiducial dust model (i.e., with $f_\mathrm{dust}=0.5$ ) with their more complex star-to-dust geometries are shown as red circles. From Fig.~\ref{fig:massiveSFgalaxies_hist}, we have estimated the average $A_\mathrm{V}$ and $A_\mathrm{V}\!-\!A_\mathrm{J}$ that are required for TNG100 to reproduce the UVJ colors of massive, star-forming galaxies in JWST/NIRCam. These values are indicated by the blue star marker in Fig.~\ref{fig:AV_AJ_relation}.

We find that the $A_\mathrm{V}$ values for the TNG100 galaxies are clustered at low values (median $A_\mathrm{V}$ of 0.35\,mag) which fall short of the attenuation required to reproduce the observational data. Using our fiducial dust radiative transfer setup, it is challenging to increase the $A_\mathrm{V}$ values of massive, star-forming TNG100 galaxies. We find that when setting $f_\mathrm{dust}=1$ ($f_\mathrm{dust}=5$) the median attenuation in the V-band only increases to 0.57\,mag (1.12\,mag). This is a consequence of the star-to-dust geometry, as some stellar populations in the TNG100 galaxies will always be unobscured. We have also experimented with a toy model where the dust mass distribution is scaled to the stellar mass distribution, with a constant dust-to-stellar mass ratio. With a dust-to-stellar mass ratio of $2\,\%$ (which is relatively high; see Fig.~\ref{fig:Mdust}), the median $A_\mathrm{V}$ is 2.03\,mag, close to the observed data. However, even in this toy model which reproduces the observed $A_\mathrm{V}$, the $\textit{V}-\textit{J}$ reddening ($A_\mathrm{V}\!-\!A_\mathrm{J}$) only reaches a median of 0.41\,mag.

\begin{figure*}
    \centering
    \includegraphics[width=\textwidth]{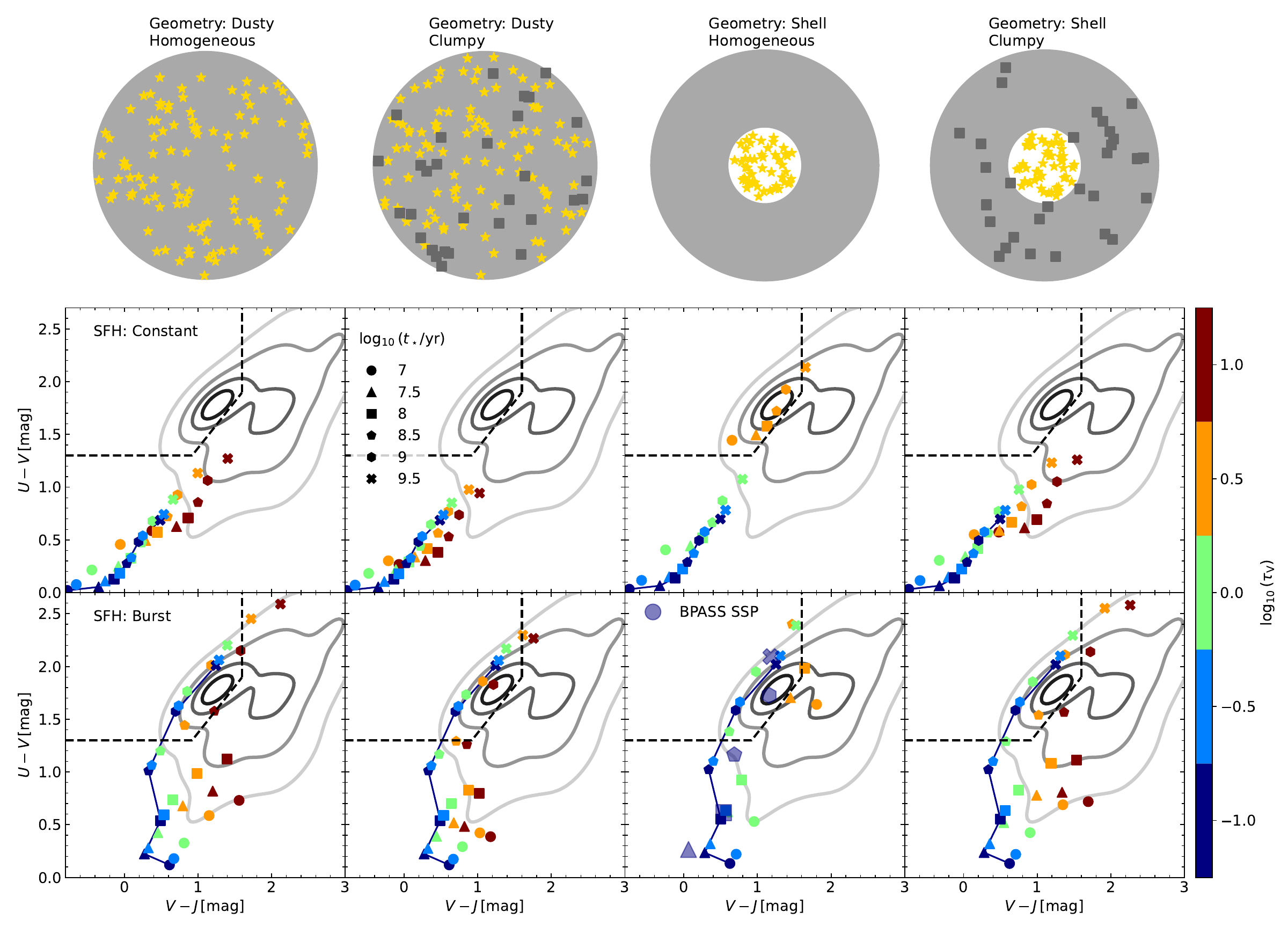}
    \caption{UVJ diagram sampling the parameter space of the DirtyGrid geometrical models. The JWST/NIRCam distribution for massive galaxies ($M_\star\geq10^{11}\,\mathrm{M}_\odot$) is also shown in each panel by the grey contours. The different columns correspond to different DirtyGrid geometries, which are schematically depicted at the top of each column. The upper panels show the results using a constant SFH, while the lower panels correspond to a single-burst SFH. Stellar ages and dust optical depths of the DirtyGrid models are indicated by the various markers and colors, respectively. We keep the stellar metallicity fixed at a metal mass fraction of 0.02. Furthermore, we use a constant stellar mass of $10^9\,\mathrm{M}_\odot$ and a fixed Milky-Way-like dust extinction curve. As reference, we also show the dust-free colors of a BPASS simple stellar population (SSP) with a metal mass fraction of 0.02 in one panel. We find that young stellar populations with a single-burst SFH and a homogeneous shell geometry (lower row, third column) can match the UVJ colors of massive, star-forming JWST/NIRCam galaxies at intermediate V-band optical depths ($\tau_\mathrm{V}\sim10^{0.5}$).}
    \label{fig:DIRTY}
\end{figure*}

Irrespective of how the dust is distributed, we generally find a flat attenuation-reddening relation for TNG100 galaxies for $A_\mathrm{V}\gtrsim0.5\,\mathrm{mag}$, while at lower $A_\mathrm{V}$ the relation tightly follows the dust screen model (see Fig.~\ref{fig:AV_AJ_relation}). Such a relation, namely, a flattening of the attenuation curve at higher dust optical depths, has been predicted in theoretical studies (\citealt{Witt2000}; \citealt{Pierini2004}; \citealt{Inoue2005}) as a signature of mixed stellar and dust distributions (\citealt{Charlot2000}; \citealt{Calzetti2001}; \citealt{Chevallard2013}). This relation was later confirmed in observations (\citealt{Salmon2016}; \citealt{Salim2018}) and cosmological, hydrodynamical simulations (\citealt{Narayanan2018}; \citealt{Shen2020}; \citealt{Trayford2020}).
 
These findings suggest that obtaining TNG100 galaxies with simultaneously high $A_\mathrm{V}$ and $A_\mathrm{V}\!-\!A_\mathrm{J}$, which is required to reproduce the very dust-reddened massive JWST/NIRCam galaxies, is challenging. However, there is a lever to modify the attenuation-reddening relation: the star-to-dust geometry. As an example, theoretical studies using simplified analytical geometries indicate that clumpier star-to-dust geometries lead to a flattening of the attenuation curve (\citealp{Witt1996,Witt2000}; \citealt{Gordon1997}). For TNG100, however, varying the star-to-dust geometry is not straightforward as the ISM and stellar distributions are direct predictions of the simulation. Hence, we now consider simplified analytical geometries in Sect.~\ref{sec:DIRTY toy model} to gain insight into which geometries are able to reproduce the observed dust reddening.

\subsection{Toy model 1: Simple geometrical models with DirtyGrid}\label{sec:DIRTY toy model}

Simplified analytical geometries lack the realistic structures of galaxies in cosmological simulations, but they are useful to understand the drivers of dust attenuation curves (\citealp{Witt1996,Witt2000}; \citealt{Gordon1997}). In this section we do not use any TNG100 simulation data but consider analytical geometries from the DirtyGrid library of toy models (\citealt{Law2018}). DirtyGrid spans an eight-dimensional (8D) parameter space with $\approx6.5\times10^6$ dust radiative transfer models that are run with the DIRTY code (\citealt{Gordon2001}; \citealt{Misselt2001}). The output of these models are global broadband fluxes in 26 bands from the FUV to the FIR, including the Johnson UVJ bands. In DirtyGrid, three classes of spherically symmetric geometries are considered (as defined in \citealt{Witt2000}): `dusty' (uniform stellar and dust distributions), `shell' (stars extend only to 0.3 times the system radius and are surrounded by dust), and `cloudy' (the stellar distribution contains a dusty core that extends up to 0.69 times the system radius). In all three geometries, the stars and dust are distributed uniformly within their radial bounds. In addition to these homogeneous models, all geometries also feature a clumpy version where dust clumps with a filling factor of 15\,\% and a density contrast of 100 are used.

Beyond these geometrical parameters, additional parameters describe the star-formation history (which is either constant or a single burst), stellar age, stellar metallicity, stellar mass, dust extinction curve, and dust optical depth (parametrized as spherically averaged V-band optical depth, $\tau_\mathrm{V}$). We show the UVJ colors of the DirtyGrid toy models in Fig.~\ref{fig:DIRTY}, with underlying grey contours indicating the distribution of the massive ($M_\star\geq10^{11}\,\mathrm{M}_\odot$) JWST/NIRCam galaxies in the UVJ diagram. The four different columns correspond to different geometries schematically depicted at the top of each column, we refrain from showing the cloudy geometry because it exhibits the weakest dust reddening. The two rows show the results for either a constant star-formation history (SFH, upper panels) or for a single burst (lower panels). The different markers indicate stellar age (ranging from $10^7\,\mathrm{yr}$ to $10^{9.5}\,\mathrm{yr}$), the colors correspond to the dust optical depth in the V-band (ranging from 0.1 to 10). An increase in stellar metallicity leads to redder SSP colors due to lower effective temperatures and stronger spectroscopic absorption features (\citealt{Conroy2013}), we find that this affects $\textit{V}-\textit{J}$ stronger than $\textit{U}-\textit{V}$. We keep the stellar metallicity in the DirtyGrid models fixed to a metal mass fraction of 0.02 since the next-highest value in the parameter space (0.1) is unrealistically high for star-forming galaxies at cosmic noon. Since the stellar mass and dust extinction parameters have no significant impact on the UVJ colors, we use a fixed Milky Way-like extinction curve and a stellar mass of $10^9\,\mathrm{M}_\odot$.

To interpret Fig.~\ref{fig:DIRTY}, it is instructive to first consider the toy models with the lowest V-band attenuation which are almost dust-free, namely, the connected dark blue markers. These markers indicate the change in UVJ colors purely due to stellar evolution from 10\,Myr to $\approx3\,\mathrm{Gyr}$. These stellar population age tracks are the same for all panels in one row. While the evolution is relatively mild for the constant SFH (upper panels), the single-burst models (lower panels) rapidly become red in $\textit{U}-\textit{V}$ color. We remark that \citet{Law2018} used the PEGASE.2 (\citealp{Fioc1997,Fioc1999}) stellar population templates to construct the DirtyGrid library. As a reference, we have also included our fiducial BPASS templates in one of the panels, using the same ages and metal mass fraction of 0.02 as used for DirtyGrid. We find moderate differences in $\textit{V}-\textit{J}$ color between the BPASS simple stellar populations and the DirtyGrid burst models with the lowest dust content, ranging from -0.2 to 0.4\,mag.

Increasing the dust optical depth reddens the DirtyGrid models, with the magnitude and direction of the reddening vector depending on the geometry. We find that a homogeneous dust shell (third column) is most effective in reddening the stars. This happens because that model mimics a dust screen, where the reddening scales linearly with the dust optical depth (as visualized by the straight line in Fig.~\ref{fig:AV_AJ_relation}). This leads to the models with the highest optical depths (dark red markers) being out of scale in the third column. On the other hand, a mixed star-to-dust geometry (the `dusty' geometry) with a clumpy dust medium (second column) is least effective. For the constant-SFH models (upper row in Fig.~\ref{fig:DIRTY}), only the homogeneous shell models are compatible with the star-forming JWST/NIRCam galaxies, although the modeled $\textit{V}-\textit{J}$ colors are still too blue (which could be remedied by a higher metallicity of the stellar populations, or different stellar population templates).

For the single-burst SFHs (lower row in Fig.~\ref{fig:DIRTY}), the homogeneous dusty and clumpy shell geometries (first and fourth columns) reach high $\textit{U}-\textit{V}$ for stellar ages above $\approx300\,\mathrm{Gyr}$, but fall short of the required $\textit{V}-\textit{J}$ color. This could be remedied again by a higher stellar metallicity or different stellar population templates, or alternatively by even higher dust optical depths. However, we note that in any case the dust optical depth in this geometry needs to be very high ($\tau_\mathrm{V}\gtrsim10$), which seems unrealistic given the EAZY-derived attenuation for massive, star-forming (selected through the UVJ demarcation line) JWST/NIRCam galaxies of $A_\mathrm{V}=2.7^{+0.8}_{-0.8}\,\mathrm{mag}$ (quoting the median and 16th-84th percentiles). The clumpy dusty geometry (second column) exhibits a similar structure, with the key difference that the dust reddening saturates at the highest optical depths such that it seems unlikely that this class of models reaches the $\textit{V}-\textit{J}$ colors observed with JWST/NIRCam. Lastly, we find that the homogeneous shell models (third column) are able to reproduce the sequence of star-forming galaxies in the observed UVJ diagram for intermediate dust optical depths ($\tau_\mathrm{V}\sim3$) and young stellar ages ($t_\star\sim10\,\mathrm{Myr}$).

We also note that the distribution of JWST/NIRCam galaxies in the quiescent region of the UVJ diagram (the top left corner) is well reproduced by burst-like SFHs with low dust content and stellar ages of $\approx1-3\,\mathrm{Gyr}$ (blue hexagons and crosses in the lower panels of Fig.~\ref{fig:DIRTY}). Given that local quiescent galaxies have traditionally been modeled as simple stellar populations (SSPs, e.g., \citealt{Worthey1994}; \citealt{Buzzoni1995}; \citealt{Thomas2003}; \citealt{Sanchez-Blazquez2006b}) we find that these analytical geometries could be realistic representations of massive quiescent galaxies at cosmic noon in the real Universe. The cosmological simulation also aligns with these results, as we find that massive, quiescent TNG100 galaxies have mass-weighted ages of $1.51^{+0.33}_{-0.41}\,\mathrm{Gyr}$ (we quote the median and 16th-84th percentile range).

On the other hand, we cannot model the star-forming galaxies by a single-burst SFH. The constant SFH could be appropriate for star-forming galaxies, but the only DIRTY geometry that potentially yields realistic UVJ colors for these galaxies is a homogeneous dust shell around the entire galaxy (third panel in the upper row of Fig.~\ref{fig:DIRTY}), which is not a physically plausible configuration. Instead, we follow \citet{Law2021} who fit the FUV-FIR SEDs of local star-forming galaxies using two or more components from the DirtyGrid parameter space, assuming that the different components do not interact. Put another way, a star-forming galaxy model in \citet{Law2021} is composed of multiple, non-interacting building blocks, which are then linearly combined (with the proper luminosity weights) to generate the SED of the full galaxy. In our case where we only consider the UVJ bands we do not have enough information to retrieve the best-fitting combination of DirtyGrid models for the massive, star-forming JWST/NIRCam galaxies reliably. To fit the JWST/NIRCam data, we can only conclude that local attenuation of SSPs (i.e., burst-like SFHs) at solar or super-solar metallicities is required. This corresponds to a superposition of multiple shell geometries that are not volume-filling.

\subsection{Toy model 2: TNG100 galaxies with analytical two-component dust attenuation}\label{sec:TNG100 toy model}

\begin{figure}
    \centering
    \includegraphics[width=\columnwidth]{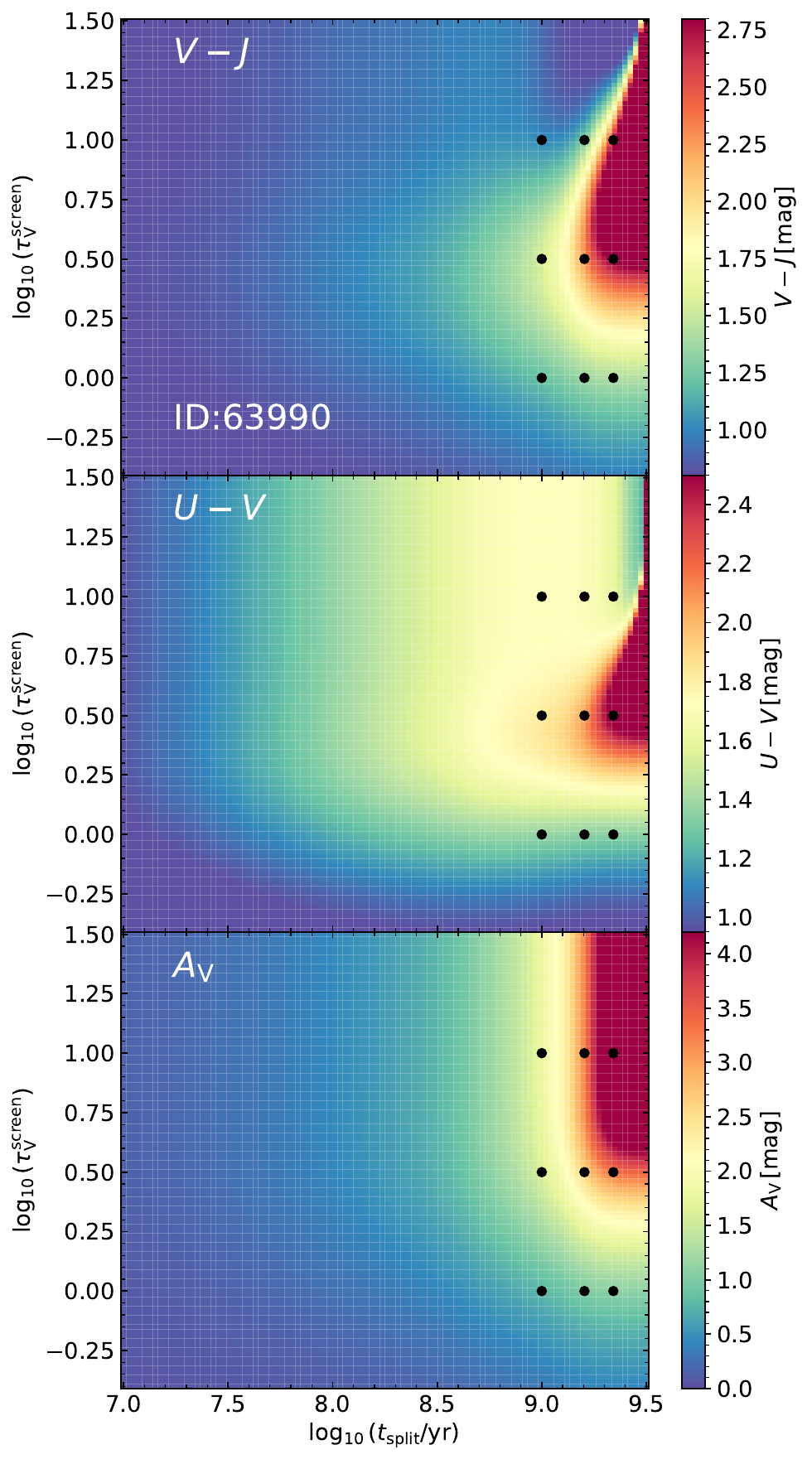}
    \caption{Application of the two-component dust attenuation toy model to a single TNG100 galaxy (subhalo ID: 63990, $M_\star\approx10^{11.2}\,\mathrm{M}_\odot$, $\mathrm{SFR}\approx140\,\mathrm{M}_\odot\mathrm{yr}^{-1}$). Shown are the $\textit{V}-\textit{J}$ color (upper panel), $\textit{U}-\textit{V}$ color (center panel), and V-band attenuation (lower panel) as a function of the two parameters of the toy model: $t_\mathrm{split}$, which denotes the age up to which stellar populations are attenuated, and $\tau_\mathrm{V}^\mathrm{screen}$, which denotes the V-band optical depth of the dust screens. The black dots mark the ($t_\mathrm{split}$, $\tau_\mathrm{V}^\mathrm{screen}$) combinations that we selected to visualize in Fig.~\ref{fig:TNG100toyModel}. To reach the very red $\textit{V}-\textit{J}$ colors observed for massive, star-forming galaxies, the stellar populations of this TNG100 galaxy need to be dust-obscured at least up to $t_\star\sim1\,\mathrm{Gyr}$.}
    \label{fig:TNG100toyModel_singleGalaxy}
\end{figure}

\begin{figure*}
    \centering
    \includegraphics[width=\textwidth]{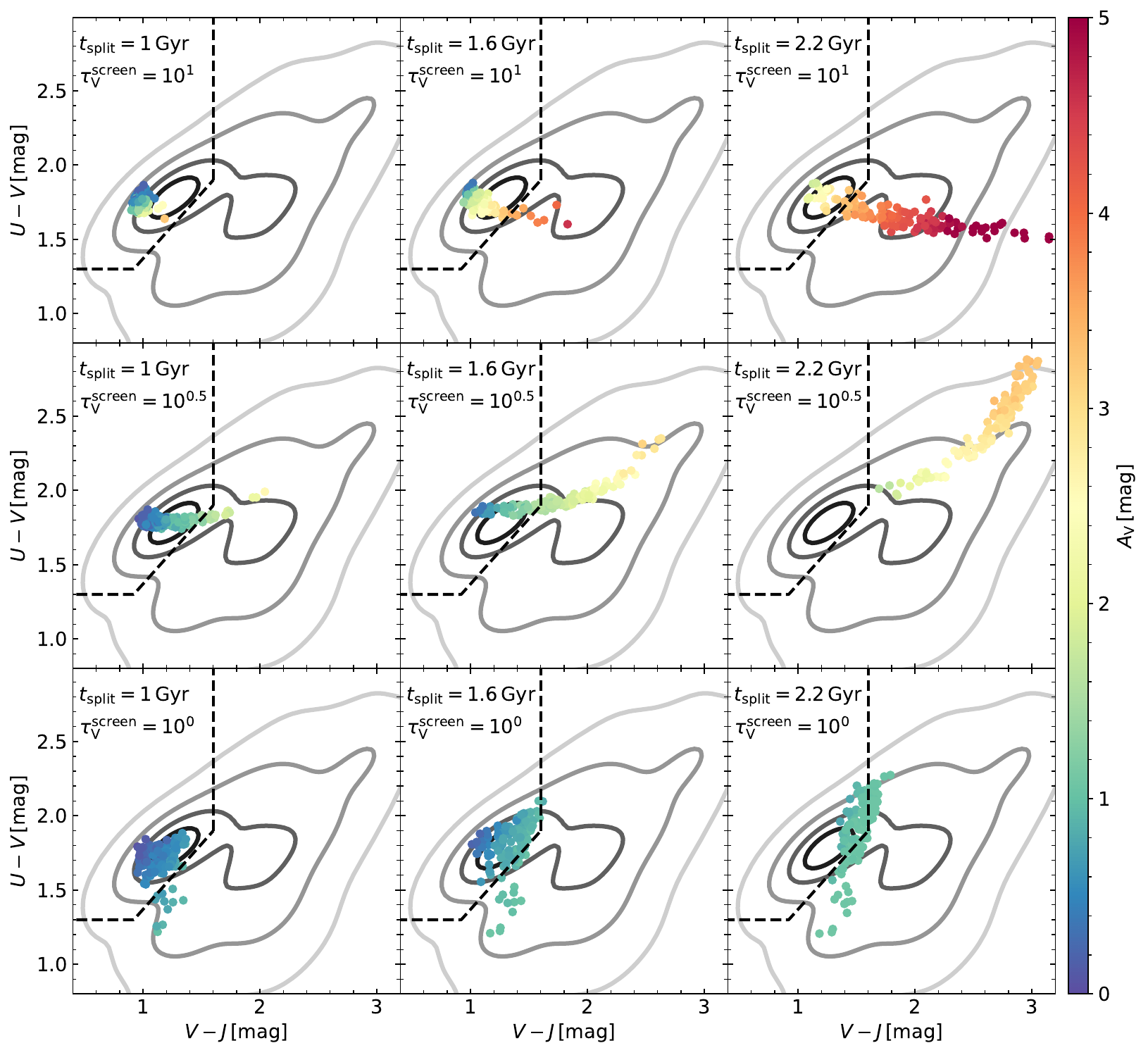}
    \caption{Results of our toy model for TNG100 galaxies with two-component dust attenuation. We show the UVJ diagram of massive, star-forming TNG100 galaxies for various combinations of the free parameters $t_\mathrm{split}$ and $\tau_\mathrm{V}^\mathrm{screen}$ in our toy model. The color-coding corresponds to the V-band attenuation for the TNG100 galaxies. We show the observational JWST/NIRCam data for massive ($M_\star>10^{11}\,\mathrm{M}_\odot$) galaxies as grey contours. The toy model in the center of the figure with dust screns of intermediate optical depths ($\tau_\mathrm{V}^\mathrm{screen}\approx3.2$) covering all stellar populations up to an age of $t_\mathrm{split}=1.6\,\mathrm{Gyr}$ provides the best match to the observational constraints.}
    \label{fig:TNG100toyModel}
\end{figure*}

Based on the insight from Sect.~\ref{sec:DIRTY toy model} that the JWST/NIRCam UVJ colors are reproducible by combining multiple SSPs, all with their own individual dust screen, we now construct such a toy model for the TNG100 galaxies. Specifically, we employ a simple two-component model where we attenuate the light of all star particles with ages below a specific threshold $t_\mathrm{split}$ with a dust screen of fixed V-band optical depth $\tau_\mathrm{V}^\mathrm{screen}$ (the older star particles are left unattenuated). Since we assume that the galaxy is composed of isolated (potentially dust-attenuated) star particles, we can simply sum the contributions of all particles to obtain the galaxy V-band flux $F_\mathrm{V}$:

\begin{equation}\label{eq:TwoComponentAttenuation}
    F_\mathrm{V} = \sum_{t_\star>t_\mathrm{split}}F_\mathrm{V,\star}+\sum_{t_\star<t_\mathrm{split}}F_\mathrm{V,\star}\mathrm{e}^{-\tau_\mathrm{V}^\mathrm{screen}},
\end{equation}
where $t_\star$ denotes stellar age and $F_\mathrm{V,\star}$ the dust-free V-band flux of a star particle. We use the BPASS templates to compute the dust-free fluxes for all star particles. The flux in the U and J bands can be obtained similarly, noting that the screen optical depths are related as follows:

\begin{equation}\label{eq:tau}
    \tau_\mathrm{U}^\mathrm{screen}=1.661\cdot\tau_\mathrm{V}^\mathrm{screen},\\
    \tau_\mathrm{J}^\mathrm{screen}=0.336\cdot\tau_\mathrm{V}^\mathrm{screen}.
\end{equation}
The numerical factors in Eq.~\ref{eq:tau} are based on the THEMIS extinction curve. With Eq.~\ref{eq:TwoComponentAttenuation}, we can analytically calculate the fluxes and attenuations in the UVJ bands without running expensive dust radiative transfer calculations, that is, without the use of SKIRT.

To visualize the effect of this two-component dust attenuation toy model, we show the application of Eq.~\ref{eq:TwoComponentAttenuation} to a single massive star-forming TNG100 galaxy (subhalo ID: 63990, $\log_{10}(M_\star/\mathrm{M}_\odot)\approx11.2$, $\mathrm{SFR}\approx140\,\mathrm{M}_\odot\mathrm{yr}^{-1}$) in Fig.~\ref{fig:TNG100toyModel_singleGalaxy}. The $\textit{V}-\textit{J}$ (upper panel) and $\textit{U}-\textit{V}$ colors (center panel) are non-trivial functions of the two free parameters of this toy model ($t_\mathrm{split}$ and $\tau_\mathrm{V}^\mathrm{screen}$). For $t_\mathrm{split}\gtrsim1\,\mathrm{Gyr}$, when monotonously increasing $\tau_\mathrm{V}^\mathrm{screen}$, the galaxy colors become redder up to $\tau_\mathrm{V}^\mathrm{screen}\sim3$. At higher optical depths the colors become bluer, because the reddened stellar populations are so dust-obscured that they do not contribute anymore to the global fluxes. The lower panel of Fig.~\ref{fig:TNG100toyModel_singleGalaxy} shows the global V-band attenuation, which is a monotonously increasing function of $t_\mathrm{split}$ and $\tau_\mathrm{V}^\mathrm{screen}$.

Importantly, the JWST/NIRCam data provides constraints for all three galaxy properties visualized in Fig.~\ref{fig:TNG100toyModel_singleGalaxy}. We show how the entire population of massive ($M_\star>10^{11}\,\mathrm{M}_\odot$), star-forming ($\mathrm{sSFR}>10^{-10.6}\,\mathrm{yr}^{-1}$) TNG100 galaxies ($N_\mathrm{gal}=131$) compares to the JWST/NIRCam constraints when applying the two-component dust attenuation toy model in Fig.~\ref{fig:TNG100toyModel}. All nine panels show various ($t_\mathrm{split}$, $\tau_\mathrm{V}^\mathrm{screen}$) parameter combinations. The `locations' of these combinations in $t_\mathrm{split}-\tau_\mathrm{V}^\mathrm{screen}$-space are marked in Fig.~\ref{fig:TNG100toyModel_singleGalaxy} by the black dots. Each panel in Fig.~\ref{fig:TNG100toyModel} shows the UVJ diagram of the TNG100 galaxies, color-coded by their $A_\mathrm{V}$ values. Given the V-band magnitudes in the third panel of Fig.~\ref{fig:massiveSFgalaxies_hist}, we expect $A_\mathrm{V}$ values of $\approx2\,\mathrm{mag}$. This also aligns with the EAZY-derived attenuation for massive, star-forming JWST/NIRCam galaxies of $A_\mathrm{V}=2.7^{+0.8}_{-0.8}\,\mathrm{mag}$ .

The toy model parameter combinations in Fig.~\ref{fig:TNG100toyModel} are chosen such that the best match to the JWST/NIRCam data is achieved by the central panel with $t_\mathrm{split}=1.6\,\mathrm{Gyr}$ and $\tau_\mathrm{V}^\mathrm{screen}\approx3.2$. Almost all other parameter combinations fall short of the required $\textit{V}-\textit{J}$ reddening, except for the toy models with $t_\mathrm{split}=2.2\,\mathrm{Gyr}$ and $\tau_\mathrm{V}^\mathrm{screen}\gtrsim3$ (upper and center right panels). In the center right panel, the reddening in both colors is overly effective for most TNG100 galaxies. The toy model in the top right panel yields realistic colors on average, but the V-band attenuation is significantly too strong. Hence, we expect that the best-fitting toy model is bound by the $t_\mathrm{split}-\tau_\mathrm{V}^\mathrm{screen}$-parameter space shown in Fig.~\ref{fig:TNG100toyModel}.

Most importantly, Fig.~\ref{fig:TNG100toyModel} means that in order to reach the very red $\textit{V}-\textit{J}$ colors that are observed in massive, star-forming galaxies, we need significant dust reddening (in the form of dust screens in this toy model) not only for star-forming regions ($t_\star\lesssim30\,\mathrm{Myr}$) but for all stellar populations up to ages of $\gtrsim1\,\mathrm{Gyr}$. For the massive, star-forming TNG100 galaxies in our sample, $33^{+22}_{-21}\,\%$ of the mass is composed of stellar population with ages below $1\,\mathrm{Gyr}$ (quoting the median and 16th-84th percentiles). At $t_\mathrm{split}=1.6\,\mathrm{Gyr}$, the mass fraction of obscured stellar populations rises to $73^{+11}_{-24}\,\%$. Hence, Fig.~\ref{fig:TNG100toyModel} demonstrates that the bulk of the stellar populations needs to be effectively dust-reddened to reproduce the JWST/NIRCam data.

This result depends on the choice of the dust extinction curve through the numerical factors in Eq.~\ref{eq:tau}. However, we note that the usage of other dust extinction curves (all based to some extent on observations of local galaxies) hardly affects the ratio $\tau_\mathrm{V}/\tau_\mathrm{J}$ (see Fig. 16 in \citealt{Hensley2023}). Additionally, all of our results depend on the intrinsic (i.e., dust-free) TNG100 fluxes, which are, in turn, determined by the stellar population properties predicted by the simulation and our choice of SED template library. To assess the impact of the stellar population properties, we also analyzed the SIMBA simulation in Sect.~\ref{sec:SIMBA}. We find that the stellar population properties are rather different comparing TNG100 and SIMBA, but this mostly affects $\textit{U}-\textit{V}$. On the other hand, the choice of SED template library significantly impacts the $\textit{V}-\textit{J}$ color while hardly affecting $\textit{U}-\textit{V}$. As shown in Appendix~\ref{sec:SKIRT variations}, BPASS (which is our fiducial template library for the evolved stellar populations) yields the reddest $\textit{V}-\textit{J}$ colors. This means that any of the other SED template libraries that we explored would require even higher $\textit{V}-\textit{J}$ dust reddening to reproduce the JWST/NIRCam data. Hence, we conclude that our main finding from Sect.~\ref{sec:DSFGs} (the bulk of the TNG100 stellar populations needs to be effectively dust-reddened) is robust under variations in the choice of cosmological simulation, SED template library, and dust extinction curve.

\section{Discussion}\label{sec:Discussion}

\begin{figure*}
    \centering
    \includegraphics[width=\textwidth]{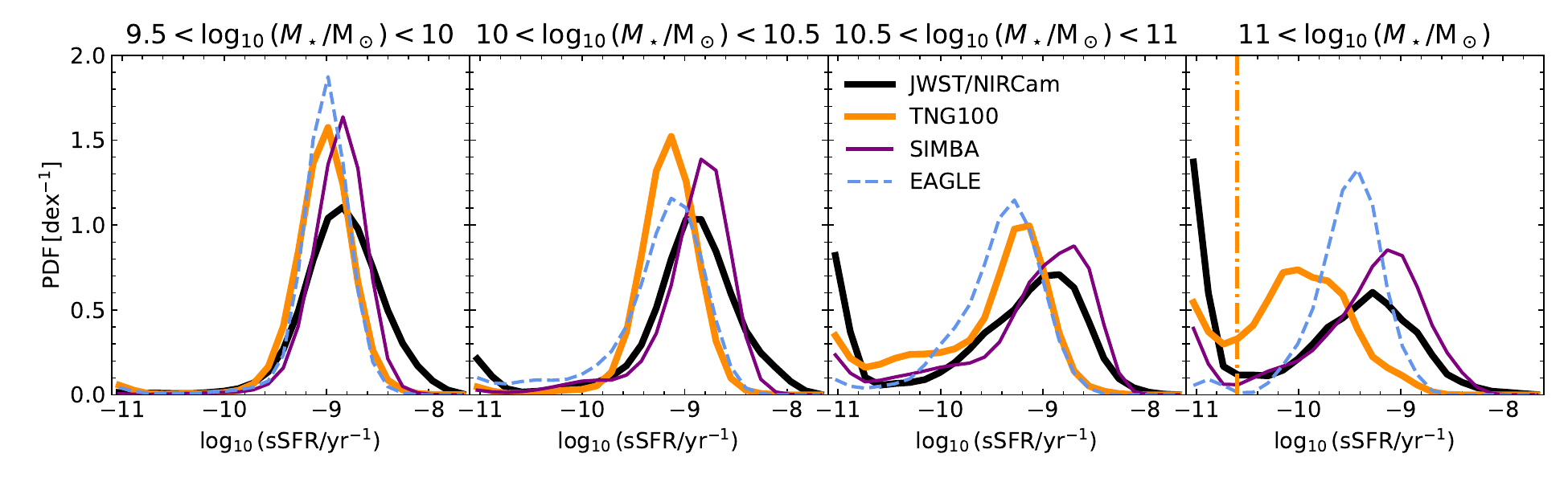}
    \caption{sSFR distribution in different stellar mass bins at redshift two, for various cosmological simulations (TNG100, SIMBA, and EAGLE) and the observational JWST/NIRCam data (derived with EAZY). The vertical orange line in the highest-mass bin marks our threshold to select massive, star-forming galaxies in TNG100. Galaxies with a sSFR below the shown $\log_{10}(\mathrm{sSFR})$ grid are assigned to the lowest sSFR bin. The galaxy main sequence in the cosmological simulations diverges at high stellar masses, but they all exhibit a population of massive, star-forming galaxies as seen in the observations.}
    \label{fig:mainSequence}
\end{figure*}

\begin{figure*}
    \centering
    \includegraphics[width=\textwidth]{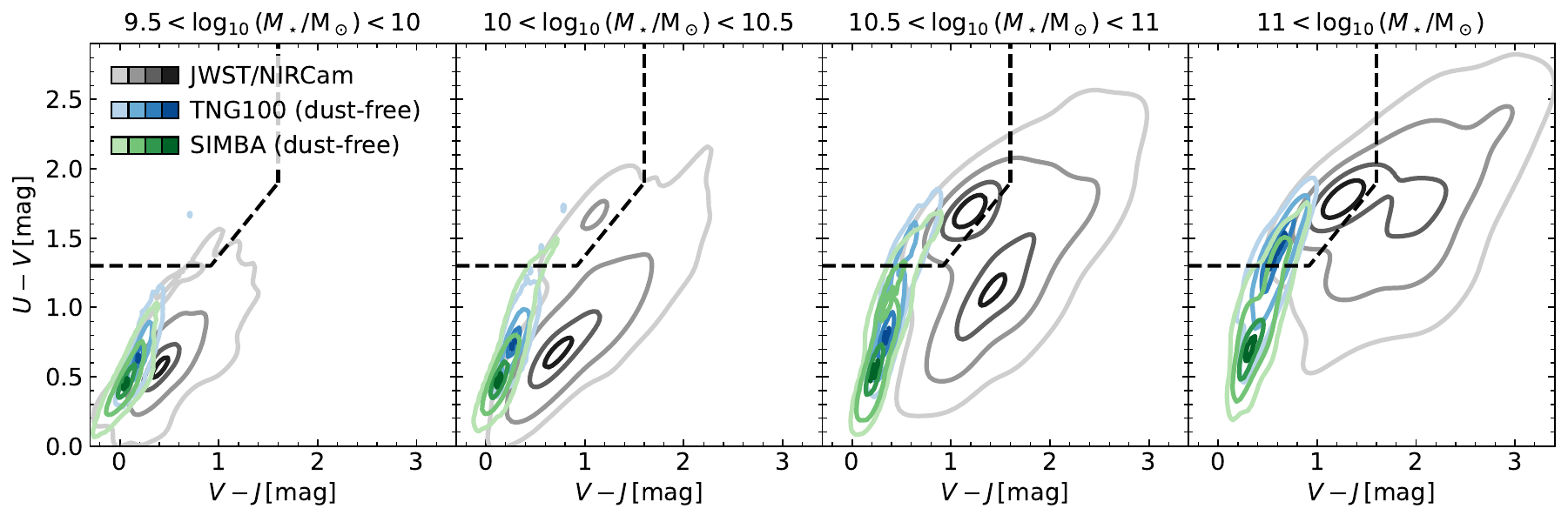}
    \caption{Comparison of the dust-free TNG100 colors (blue contours) to the results from \citet{Akins2022} for the SIMBA simulation (green contours). For both simulations, the colors are obtained with FSPS-MILES such that all differences in the UVJ distributions reflect variations in the underlying stellar populations. For massive galaxies, the SIMBA galaxies exhibit bluer $\textit{U}-\textit{V}$ colors which is due to their higher star-formation rates (Fig.~\ref{fig:mainSequence}).}
    \label{fig:UVJ_SIMBA}
\end{figure*}

\subsection{Simulation caveats}\label{sec:Simualation caveats}

All of our results depend on the properties of the simulated TNG100 galaxies, since the intrinsic (dust-free) fluxes are determined by the ages and metallicities of the stellar populations. To assess the systematic uncertainty that arises from the cosmological simulation, it is instructive to compare the stellar population properties to other simulations at $z=2$. This is shown in Fig.~\ref{fig:mainSequence}, where we plot sSFR histograms in bins of stellar mass for the TNG100, EAGLE (\citealt{Schaye2015}; \citealt{Crain2015}), and SIMBA (\citealt{Dave2019}) simulations. These simulations have comparable volumes ($10^6\,\mathrm{cMpc}^3$ for EAGLE, $1.4\times10^6\,\mathrm{cMpc}^3$ for TNG100, and $3.2\times10^6\,\mathrm{cMpc}^3$ for SIMBA), while the baryonic mass resolution of SIMBA ($1.8\times10^7\,\mathrm{M}_\odot$) is somewhat lower than the one of TNG100 ($1.4\times10^6\,\mathrm{M}_\odot$) or EAGLE ($1.8\times10^6\,\mathrm{M}_\odot$).

To guide the eye, we also show the sSFR distributions for the observational dataset in Fig.~\ref{fig:mainSequence}, as these have been measured with EAZY and included in the JWST/NIRCam catalog. We emphasize that Fig.~\ref{fig:mainSequence} is the only instance where we use derived quantities for the observed galaxies other than redshift and stellar mass. We intentionally keep the main comparison between cosmological simulations and JWST/NIRCam data limited to `observable' properties like colors and magnitudes, as the inference of more complex derived quantities via SED fitting comes with significant systematic uncertainties due to simplifying assumptions about star-formation histories and star-to-dust geometries (\citealt{Pacifici2023}). Hence, Fig.~\ref{fig:mainSequence} serves as a sketch to test whether the galaxy populations at cosmic noon in the cosmological simulations and in the observations are broadly compatible, but we refrain from drawing more stringent conclusions based on this comparison. 

At low stellar masses, all cosmological simulations display similar sSFR distributions with a single peak corresponding to the main sequence of star-forming galaxies, in good agreement with the JWST/NIRCam data. At higher stellar masses, the quiescent galaxy population emerges and the star-forming sequences in the simulations increasingly diverge. In the highest stellar mass bin, the modes of the TNG100 and SIMBA sSFR distributions differ by $\approx1.2\,\mathrm{dex}$.

To investigate how the large uncertainty in star-formation rates for the massive galaxies in the simulations impacts our results, it is instructive to inspect the dust-free TNG100 and SIMBA UVJ colors shown in Fig.~\ref{fig:UVJ_SIMBA}. These are calculated with the same SED templates (FSPS-MILES\footnote{We use the shorthand FSPS-MILES for the SED template library calculated with FSPS (\citealt{Conroy2009}; \citealt{Conroy2010}) using the MIST isochrones (\citealp{Paxton2011,Paxton2013,Paxton2015}) and the MILES stellar library (\citealt{Sanchez-Blazquez2006a}).}) to ensure that the differences in intrinsic UVJ colors are purely due to different stellar population properties in SIMBA and TNG100. We find that the discrepancy in the dust-free colors increases with stellar mass, reaching a difference of 0.17\,mag (0.49\,mag) in the median $\textit{V}-\textit{J}$ ($\textit{U}-\textit{V}$) colors for the highest stellar mass bin. This is a consequence of the SFR differences in the SIMBA and TNG100 galaxy populations (see Fig.~\ref{fig:mainSequence}), with the SIMBA galaxies being more star-forming at $z=2$ and hence bluer (especially in $\textit{U}-\textit{V}$). 

We have based the discussion in Sect.~\ref{sec:DSFGs} mostly on $\textit{V}-\textit{J}$ because the tension with observational data is much more severe in this color than in $\textit{U}-\textit{V}$ (see Fig.~\ref{fig:massiveSFgalaxies_hist}), but also because the intrinsic (i.e., dust-free) $\textit{V}-\textit{J}$ is less sensitive to the underlying cosmological simulation. For instance, the relatively small difference in intrinsic $\textit{V}-\textit{J}$ color between TNG100 and SIMBA means that the main result from Sect.~\ref{sec:DSFGs} (a steep attenuation-reddening relation is needed to reproduce the observational data, which can only be achieved by effectively reddening the bulk of the stellar populations) is robust against variations in the cosmological simulation. On the other hand, any conclusions drawn for $\textit{U}-\textit{V}$ are subject to this systematic uncertainty.

This has profound implications on our results from Sect.~\ref{sec:MassResolvedUVJ}: Starting from the dust-free TNG100 fluxes, the dust attenuation vector needs to be relatively flat in order to reproduce the massive, star-forming JWST/NIRCam data in the UVJ diagram. This is partially mitigated by our choice of the BPASS SED templates which lead to the reddest intrinsic $\textit{V}-\textit{J}$ colors, but it still seems challenging to reproduce the observed $\textit{V}-\textit{J}$ color distribution while not reddening $\textit{U}-\textit{V}$ too much. We also encountered this problem in Sect.~\ref{sec:TNG100 toy model}, where we found that a screen-like attenuation for TNG100 galaxies can match the observed $\textit{V}-\textit{J}$ colors but yields on average too blue $\textit{U}-\textit{V}$ colors. If the massive TNG100 galaxies at cosmic noon were more star-forming (which is also required by their sSFR distributions, Fig.~\ref{fig:mainSequence}), their stellar populations would be younger on average leading to bluer $\textit{U}-\textit{V}$ colors. This would give additional `wiggle room' for the dust attenuation to sufficiently redden the $\textit{V}-\textit{J}$ color of the simulated massive, star-forming galaxies.

Overall, we conclude that the choice of cosmological simulation does not affect our main finding that the dust radiative transfer methods explored here cannot reproduce the mass-resolved UVJ diagram seen in observations at cosmic noon, unless we employ screen-like attenuation (Sect.~\ref{sec:TNG100 toy model}). Still, with a cosmological simulation that forms stars more intensely in massive galaxies at cosmic noon leading to bluer intrinsic $\textit{U}-\textit{V}$ colors, it would be easier to find a dust attenuation recipe which reproduces the observational data.

\subsection{Comparison to SIMBA (Akins et al. 2022)}\label{sec:SIMBA}

The cosmological SIMBA simulation provides a particularly interesting dataset to compare to, as SIMBA includes an on-the-fly subgrid dust model (\citealt{Dave2019}; \citealt{Li2019}), which circumvents the need to assign dust to gas cells in a post-processing step (as is required for TNG100). The UVJ diagram for SIMBA at cosmic noon has been studied by \citet{Akins2022} using the dust radiative transfer code POWDERDAY (\citealt{Narayanan2021}). This is a similar methodology as the one we adopt here, but numerous differences related to the choice of SED templates (\citealt{Akins2022} adopt the FSPS-MILES templates), treatment of star-forming regions, and dust optical properties exist. We compare their dust-attenuated UVJ colors to our results in the upper panels of Fig.~\ref{fig:UVJ_simComparison}.

Given that the SIMBA galaxies exhibit intrinsically bluer colors (especially in $\textit{U}-\textit{V}$; see Fig.~\ref{fig:UVJ_SIMBA}), we note that the reddening in the UVJ colors is significantly more effective in SIMBA (especially for the massive galaxies). Selecting only massive ($M_\star>10^{11}\,\mathrm{M}_\odot$), star-forming ($\mathrm{sSFR}>10^{-10.6}\,\mathrm{yr}^{-1}$) galaxies, we find the following median reddening and attenuation values for SIMBA: $A_\mathrm{V}-A_\mathrm{J}\approx0.69\,\mathrm{mag}$, $A_\mathrm{U}-A_\mathrm{V}\approx0.73\,\mathrm{mag}$, $A_\mathrm{V}\approx2.3\,\mathrm{mag}$. These values are significantly higher than the corresponding attenuation properties for TNG100 (see also Fig.~\ref{fig:massiveSFgalaxies_hist}): $A_\mathrm{V}-A_\mathrm{J}\approx0.25\,\mathrm{mag}$, $A_\mathrm{U}-A_\mathrm{V}\approx0.29\,\mathrm{mag}$, $A_\mathrm{V}\approx0.35\,\mathrm{mag}$. The high V-band attenuation in SIMBA makes the massive, star-forming SIMBA galaxies consistent with the JWST/NIRCam data in $M_\mathrm{V}$ (V-band magnitude), which is not the case for TNG100 (third panel in Fig.~\ref{fig:massiveSFgalaxies_hist}). On the other hand, also the SIMBA galaxies exhibit a flat attenuation-reddening relation ($A_\mathrm{V}$ versus $A_\mathrm{V}-A_\mathrm{J}$, Fig.~\ref{fig:AV_AJ_relation}), which causes insufficient dust reddening in $\textit{V}-\textit{J}$ to reproduce the observations.

The significantly stronger dust attenuation in SIMBA compared to TNG100 is particularly interesting when comparing the specific dust masses between the two simulations. For massive, star-forming galaxies the median specific dust masses are slightly higher in TNG100 (by $\approx0.1\,\mathrm{dex}$), while the 16th-84th percentile range is higher in SIMBA (by $\approx0.2\,\mathrm{dex}$). Given these remarkably similar dust masses, the stark differences in dust attenuation between SIMBA and TNG100 must arise from different star-to-dust geometries. We remark that this is unlikely due to the different resolutions of SIMBA and TNG100: We have also calculated the UVJ colors for massive galaxies from the TNG300-1 simulation which has a comparable resolution as SIMBA, and found no significant differences in the UVJ colors between TNG100 and TNG300-1.

\subsection{Comparison to EAGLE (Camps et al. 2018)}\label{sec:EAGLE}

\begin{figure*}
    \centering
    \includegraphics[width=\textwidth]{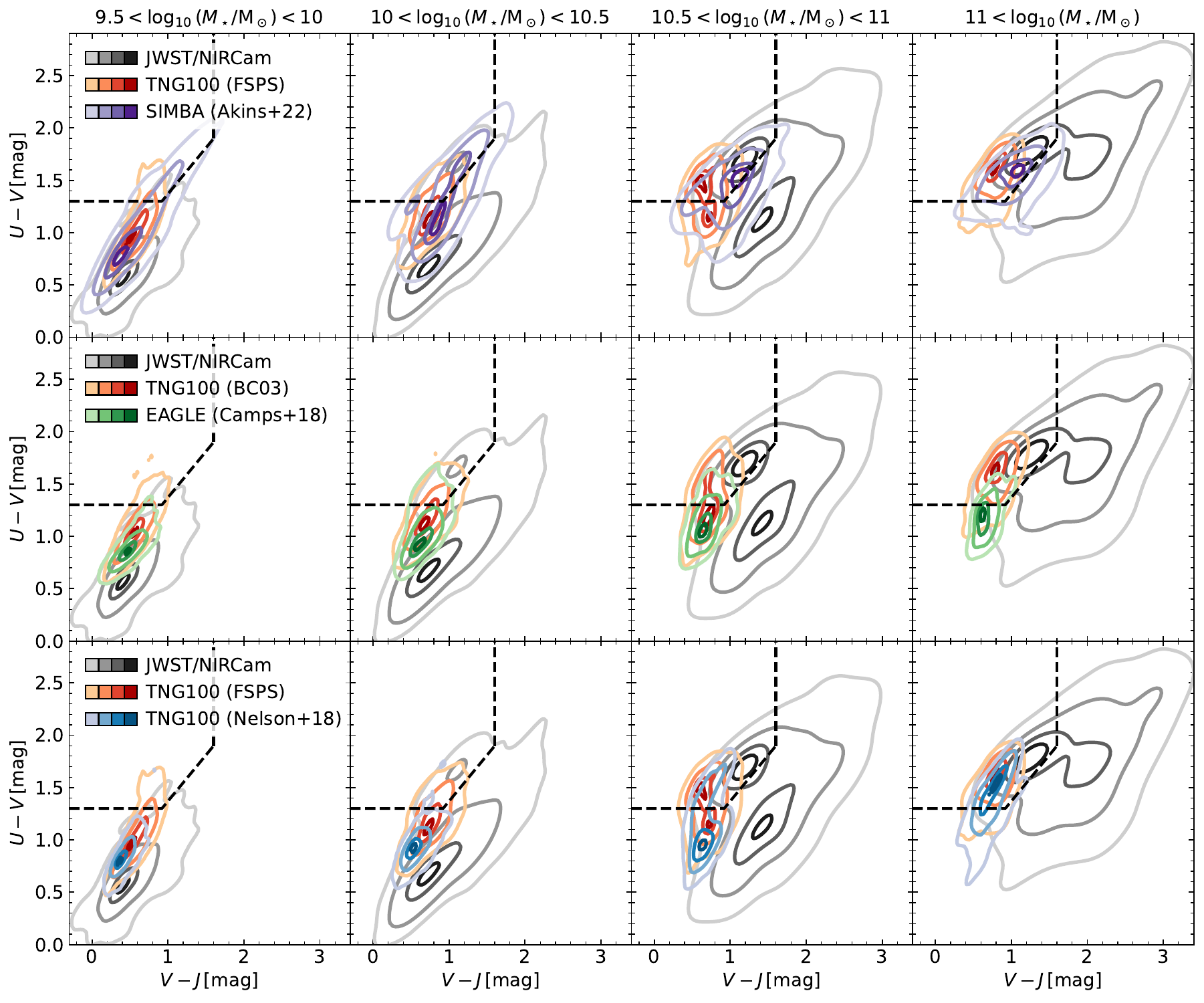}
    \caption{Comparison of dust-attenuated TNG100 UVJ colors to various simulation datasets. In all rows, our TNG100 colors (red contours) correspond to our fiducial SKIRT post-processing including dust, but varying the SED templates (either using \citealt{Bruzual2003} or FSPS-MILES) to be consistent with the simulated colors that we compare to. In the upper panels, we compare to UVJ colors from \citet{Akins2022} from the SIMBA simulation (purple contours). The center panels show the comparison to the UVJ colors from \citet{Camps2018} for the EAGLE simulation (green contours). The lower panels show a comparison to an alternative method to obtain the dust-attenuated UVJ colors for TNG100 based on raytracing (\citealt{Nelson2018}, blue contours). Differences in the models arise due to the usage of different dust attenuation methods, varying the treatment of young stellar populations, and differences in the cosmological simulation, but we note that all models fail to reproduce the massive dust-reddened galaxies seen with JWST/NIRCam.}
    \label{fig:UVJ_simComparison}
\end{figure*}

For the EAGLE cosmological simulation, \citet{Camps2018} computed publicly available\footnote{\url{http://virgodb.dur.ac.uk:8080/Eagle/Help?page=databases/fiducial_models/refl0100n1504_dustymagnitudes}} dust-attenuated fluxes for galaxies with $0\leq z\leq6$ with SKIRT. These fluxes are only available for galaxies with a well-resolved dust distribution (more than 250 dust-containing gas particles). This criterion removes 56 galaxies out of the 3568 EAGLE galaxies at $z=2$ with $M_\star\geq10^{9.5}\,\mathrm{M}_\odot$. We compare the dust-attenuated UVJ colors of the remaining EAGLE galaxies at $z=2$ to our TNG100 results in the central panels of Fig.~\ref{fig:UVJ_simComparison}, where we use the fiducial setup but replace the BPASS with the BC03 templates to be consistent with \citet{Camps2018}. Because the EAGLE database does not contain dust-free fluxes in the UVJ bands, we do not analyze the dust attenuation properties for EAGLE galaxies in detail but only compare the dust-attenuated EAGLE and TNG100 UVJ colors.

We find that the EAGLE galaxies are bluer in $\textit{U}-\textit{V}$, with the difference increasing with stellar mass. We attribute this primarily to the higher SFRs for massive EAGLE galaxies at $z=2$ (see Fig.~\ref{fig:mainSequence}), meaning they host younger (and hence bluer) stellar populations. Additional differences could arise because of different star-to-dust geometries in TNG100 and EAGLE. We use a similar method as \citet{Camps2018} to distribute dust, but the criteria to determine whether a gas particle/cell contains dust is slightly different (we use the density and temperature-dependent criterion from \citealp{Torrey2012, Torrey2019}, while \citet{Camps2018} assign dust to all gas particles that are either star-forming or have a temperature below 8000\,K). Furthermore, we use $f_\mathrm{dust}=0.5$, while \citet{Camps2018} use $f_\mathrm{dust}=0.3$, which could explain why also the low-mass EAGLE galaxies are less reddened than the TNG100 galaxies. Lastly, \citet{Camps2018} apply a resampling prescription for star-forming regions and use the MAPPINGS-III templates, while we use the TODDLERS templates for star-forming regions without resampling.

\subsection{Comparison to Nelson et al. 2018 (TNG100)}

\citet{Donnari2019} investigated the main sequence and quenched fractions of TNG100 galaxies for $0\leq z\leq2$. Of specific interest to our study is their usage of the UVJ diagram as a tool to distinguish star-forming and quiescent galaxies. \citealt{Donnari2019} noted that they do not reproduce the most dust-reddened UVJ colors that are seen in the observational CANDELS data (\citealt{Fang2018}), which is a challenge that we also encounter.

To compute the dust-attenuated UVJ colors, \citet{Donnari2019} rely on "Dust model C" from \citet{Nelson2018}. In this model, the publicly available\footnote{\url{https://www.tng-project.org/data/docs/specifications/\#sec5k}} galaxy colors are computed using FSPS-MILES templates (with Padova isochrones, as opposed to the MIST isochrones used by \citet{Akins2022} and this work) with an unresolved dust attenuation prescription following \citet{Charlot2000}. Subsequently, the star particles are attenuated by the resolved dust by computing the line-of-sight extinction using the solar neighbourhood extinction curve from \citealt{Cardelli1989}. The total (absorption plus scattering) optical depth scales with the gas-phase metallicity and neutral hydrogen column density along the line-of-sight (see Sect. 3.3 of \citealt{Nelson2018} for more details). Importantly, scattering of starlight into the line-of-sight cannot be modeled with this approach.

We compare the UVJ colors studied by \citet[based on \citealt{Nelson2018}]{Donnari2019} to our results in the lower panels of Fig.~\ref{fig:UVJ_simComparison}. To enable a more accurate comparison of the methods, we use FSPS-MILES instead of BPASS for our TNG100 results (our other settings correspond to the fiducial setup including dust). We find that the UVJ colors from \citet{Nelson2018} are slightly bluer. This difference could arise from the different treatment of young stellar populations, different dust extinction curves, variations in the dust distribution, or different methods to model scattering. Importantly, the method to compute UVJ colors for TNG100 galaxies from \citet{Nelson2018} broadly aligns with our results from more sophisticated 3D dust radiative transfer, and both methods fail to reproduce the observational data.

\subsection{Clumpy star-to-dust geometry: observational hints}

From Sect.~\ref{sec:DSFGs}, we learn that we need a specific star-to-dust geometry to make the massive TNG100 galaxies at $z\sim2$ red enough. Specifically, a promising scenario consists in a clumpy dust distribution which is localized around younger stellar populations (with ages $\lesssim1\,\mathrm{Gyr}$) and not volume-filling. Contrary to the standard approach in dust radiative transfer modeling which sees the diffuse ISM playing the most relevant role in dust reddening, for the enigmatic dusty star-forming galaxies at cosmic noon the reddening of starlight mostly occurs in local dust clouds in this toy model. A clumpy star-to-dust geometry has recently been supported by a great store of observational evidence based on sampling the stellar and the cold dust emission of DSFGs at cosmic noon (e.g., \citealp{Tadaki2020,Tadaki2023}; \citealp{Pantoni2021a,Pantoni2021b}; \citealt{Polletta2024}; \citealt{Hodge2025}).

Spatially resolved observations taken with ground-based adaptive optics facilities (e.g., SINFONI on the Very Large Telescope), HST, and (more recently) JWST revealed that DSFGs at cosmic noon are characterized by a disturbed and irregular morphology in the UV and optical rest-frame, dominated by active sites of star formation dubbed `clumps' (e.g., \citealp{ForsterSchreiber2011a,ForsterSchreiber2011b}; \citealt{Targett2013}; \citealp{Guo2015,Guo2018}; \citealp{Zanella2015,Zanella2019}; \citealt{Rujopakarn2019}; \citealt{Hodge2019}; \citealt{Hodge2020}; \citealt{Pantoni2021b}; \citealt{Kalita2024}; \citealt{Polletta2024}; \citealt{LeBail2024}). These clumps have typical stellar masses $M_\star\sim10^7-10^9\,\mathrm{M}_\odot$, $\mathrm{SFR}\sim0.1-10\,\mathrm{M}_\odot\mathrm{yr}^{-1}$, and mostly unresolved or barely resolved sizes of $\lesssim1\,\mathrm{kpc}$ up to a few kpc (although morphological fitting also reveals a smooth stellar disk component, e.g., \citealt{Targett2013}; \citealt{Rujopakarn2019}; \citealt{Hodge2019}; \citealt{Miller2022}). The observed optical emission (when dust attenuation allows it to emerge as opposed to the UV/optically dark dusty galaxies, e.g., \citealt{Shu2022}; \citealt{Barrufet2023}; \citealt{Giulietti2023}) traces the most recent star-forming regions, while the NIR emission traces the older stellar population. In this respect, \citet{Kalita2024} combined HST/ACS and JWST/NIRCam data at $0.5-4.6\,\mu\mathrm{m}$ (from the CEERS survey, \citealp{Finkelstein2017,Finkelstein2023}) to study the optical/NIR clumps up to the resolution limit ($\sim1\,\mathrm{kpc}$) in a stellar mass complete sample of star-forming galaxies at $1<z<2$. The majority of optical clumps have a NIR counterpart, which are found to follow the UVJ characteristics of the host galaxy and are thought to have a dominant role in determining the host galaxy color, its sSFR and dust attenuation. These star-forming clumps are characterized by elevated attenuation values that can reach $A_\mathrm{V}=5-7\,\mathrm{mag}$ (e.g., \citealt{Polletta2024}; \citealt{LeBail2024}), which further suggests that the high attenuation levels in DSFGs can be traced back to dust in star-forming clumps rather than to dust in the diffuse ISM.

ALMA emission from cold dust is usually observed in the most central regions of DSFGs and has typical sizes of $\lesssim1\,\mathrm{kpc}$, often limited by spatial resolution. In most cases, the FIR/sub-mm peak does not match the peak of stellar emission, which is usually observed at slightly larger radii ($\Delta r\sim1\,\mathrm{kpc}$). Hence, ALMA is thought to trace the dust-obscured and most intense star-forming region in DSFGs, whose emission is not detected at shorter wavelengths (e.g., \citealt{Hodge2020}; \citealt{Pantoni2021b}; \citealt{Hamed2023}; \citealt{Hodge2025}). In accordance with this scenario, \citet{Miller2022} who studied a subsample of 54 massive star-forming galaxies at cosmic noon selected from 3D-HST which are also present in the JWST/NIRCam CEERS survey, found the majority of star-forming galaxies in their sample to have negative gradients in both $\textit{U}-\textit{V}$ and $\textit{V}-\textit{J}$ colors (i.e., redder in the center, bluer in the outskirts). This is consistent with radially decreasing dust attenuation, with the highest levels of dust attenuation found in the central regions (see also \citealt{Shivaei2024} who reach the same conclusion based on UV-MIR photometry including JWST/MIRI).

\citet{Hamed2023} found that the compact cold dust emission, extended stellar radii, and the clumpiness typical of these galaxies call for shallow curves and double exponential attenuation laws to account for the missing photons absorbed by dust. The same result is found by \citet{Pantoni2021a} but on a much smaller sample of DSFGs at $z\sim2$ and, more recently, by \citet{Polletta2024}, for two DSFGs in their sample (using NIRCam data at rest-frame wavelength $0.4-1.2\,\mu\mathrm{m}$). The attenuation curves found by the latter works differ from most of the standard attenuation curves (e.g., \citealt{Cardelli1989}; \citealt{Calzetti2000}; \citealt{Battisti2022}), possibly implying that the dust grain size distributions and compositions in these galaxies might be different than what is commonly observed in the nearby Universe or other galaxy types.

Finally, \citet{Lorenz2023}, who investigated the dust attenuation and its dependence on viewing angle in a statistical sample of 308 DSFGs at $1.3<z<2.6$ from the MOSFIRE Deep Evolution Field survey (\citealt{Kriek2015}), do not find any significant variation in the Balmer decrement, the attenuation in the V band, or the UV continuum spectral slope with galaxy inclination. The authors claim that this result clearly supports a geometry in which the diffuse ISM does not play a large role in attenuation, while the major contribution must be ascribed to very dusty and young star-forming clumps. The same conclusion is reached by \citet{Zhang2023} who use observational constraints from UV-NIR and FIR for star-forming galaxies at $0<z<2.5$ to construct toy models with SKIRT. They found that only clumpy star-to-dust distributions fulfill the observational constraints, similarly to what we see in this work and to what emerges from the aforementioned most recent observational studies (but noting \citealt{Zuckerman2021}, who claimed that the spread of UVJ colors is almost entirely due to inclination effects, based on their analytical dust attenuation calculations).

\section{Conclusions}\label{sec:Conclusion}

We have analyzed the UVJ color-color diagram for cosmological, hydrodynamical simulations at cosmic noon ($z\approx2$). We  performed an in-depth comparison of the mass-resolved UVJ diagram (i.e., the UVJ diagram in bins of stellar mass) to observational data from JWST/NIRCam, focusing on the enigmatic population of massive dusty star-forming galaxies. To this end, we post-processed the TNG100 simulation with the SKIRT dust radiative transfer code to generate dust-attenuated UVJ fluxes for the simulated galaxies. We also compared our results to the EAGLE and SIMBA simulations and the DirtyGrid geometrical models. The following points summarize our findings:

\begin{itemize}

\item The intrinsic (i.e., dust-free) TNG100 colors form a steep and narrow sequence in the UVJ color-color diagram (Fig.~\ref{fig:UVJ_sSFR}). The location of a galaxy along this sequence (i.e., the $\textit{U}-\textit{V}$ color) is strongly correlated  with the sSFR.

\item The addition of dust reddens the simulated galaxies and shifts them towards the top right in the UVJ diagram (Fig.~\ref{fig:UVJ_diffuseDust}). In TNG100, the amount of dust is controlled via the free dust-to-metal ratio parameter, $f_\mathrm{dust}$. We find that upon increasing the value of $f_\mathrm{dust}$, the UVJ colors of star-forming TNG100 galaxies are indeed reddened. However, this reddening saturates quickly such that an increase in $f_\mathrm{dust}$ from 0.5 to 1 affects the UVJ colors in all stellar mass bins by less than 0.05\,mag. We also note that low-mass star-forming galaxies are too red in our fiducial TNG100 models with $f_\mathrm{dust}=0.5$, meaning that a constant dust-to-metal ratio cannot reproduce the observational data.

\item We do not reproduce the UVJ colors of massive ($M_\star>10^{11}\,\mathrm{M}_\odot$), star-forming galaxies in JWST/NIRCam that are heavily dust-reddened (DSFGs). This statement is robust against variations in the cosmological simulation, SED templates, treatment of star-forming regions, and dust models. While the $\textit{U}-\textit{V}$ colors of JWST/NIRCam galaxies can be reproduced in the simulations, all dust radiative transfer models investigated, here yield ${\textit{V}}-{\textit{J}}$ colors that are significantly too blue (for our fiducial TNG100 model, the discrepancy is $\approx0.9\,\mathrm{mag})$. Additionally, we observe that the V-band attenuation is not strong enough, leading to the TNG100 galaxies being too bright by $\approx1.6\,\mathrm{mag}$ (Fig.~\ref{fig:massiveSFgalaxies_hist}).

\item To reproduce the JWST/NIRCam data for DSFGs, we find that an attenuation-reddening relation ($A_\mathrm{V}$ versus $A_\mathrm{V}\!-\!A_\mathrm{J}$) as steep as a dust screen model is required (Fig.~\ref{fig:AV_AJ_relation}). All of our dust radiative transfer models exhibit significantly flatter attenuation-reddening relations, leading to a lack of $\textit{V}-\textit{J}$ reddening for the simulated galaxies.

\item We investigate various star-dust-arrangements with simple geometrical models using DirtyGrid, a library of models with varying geometries, dust contents, and stellar populations. At solar metallicity, we find that the only models which reproduce the JWST/NIRCam data for DSFGs are homogeneous dust shells around simple stellar populations. This is the only toy model geometry, which exhibits a steep enough attenuation-reddening relation such that it yields red enough $\textit{V}-\textit{J}$ colors at realistic V-band attenuations ($A_\mathrm{V}\sim2-3\,\mathrm{mag}$).

\item Applying such a toy model of screen-like dust attenuation for individual stellar populations, we find that massive, star-forming TNG100 galaxies need dust screens around the bulk of their stellar populations ($t_\mathrm{split}\gtrsim1\,\mathrm{Gyr}$) to reproduce the observational data (Fig.~\ref{fig:TNG100toyModel}). Such a dust attenuation model comprised of "isolated" dust screens and clouds around stellar populations is vastly different from our fiducial dust radiative transfer models, which are widely used and verified at low redshift (e.g., \citealt{Gebek2024}).

\end{itemize}

We have limited this study to the analysis of only few parameters (stellar masses and rest-frame UVJ fluxes), which are relatively reliable to derive from observational photometric data. This allows a robust comparison of these quantities for statistical samples of galaxies, but in order to correctly model and understand dust attenuation specifically for massive, dusty star-forming galaxies at cosmic noon, the inclusion of additional information seems warranted. Dust attenuation can be constrained by spatially resolved JWST/NIRCam imaging (e.g., \citealt{Polletta2024}) and a comparison of stellar and nebular reddening (measurable for instance with JWST NIRSPEC) can offer insights into the differential attenuation of older versus younger stars (e.g., \citealt{Shivaei2020}, \citealt{Cooper2024}; \citealt{Lorenz2024}). Utilizing rest-frame FIR information (measurable with ALMA), we can gain additional insights into the dust content and distribution by measuring dust emission. With these constraints, we envision the potential to refine dust radiative transfer modeling at higher redshifts and to shed light on the workings of dust attenuation for the enigmatic dusty star-forming galaxies at cosmic noon.

\begin{acknowledgements}
We thank Hollis Akins for sharing their SIMBA data products and Peter Camps for insightful discussions regarding SKIRT. We also wish to acknowledge the constructive feedback from the anonymous referee.

\\

AG gratefully acknowledges financial support from the Fund
for Scientific Research Flanders (FWO-Vlaanderen, project
FWO.3F0.2021.0030.01). 

\\

This study made extensive use of the \texttt{python} programming
language, especially the \texttt{numpy} (\citealt{Harris2020}), \texttt{matplotlib} (\citealt{Hunter2007}), \texttt{astropy} (\citealp{Astropy2013,Astropy2018,Astropy2022}), and \texttt{scipy} (\citealt{Virtanen2020}) packages. We also acknowledge the use of the \texttt{TOPCAT} visualization tool (\citealt{Taylor2005}).

\\

This work made use of v2.2.1 of the Binary Population and Spectral Synthesis (BPASS) models as described in \citet{Eldridge2017} and \citet{Stanway2018}.

\\

Some of the data products presented herein were retrieved from the DAWN JWST Archive (DJA). DJA is an initiative of the Cosmic Dawn Center (DAWN), which is funded by the Danish National Research Foundation under grant DNRF140. 

\\

This work is based on observations taken by the 3D-HST Treasury Program (GO 12177 and 12328) with the NASA/ESA HST, which is operated by the Association of Universities for Research in Astronomy, Inc., under NASA contract NAS5-26555.

\\

The IllustrisTNG simulations were undertaken with compute time awarded by the Gauss Centre for Supercomputing (GCS) under GCS Large-Scale Projects GCS-ILLU and GCS-DWAR on the GCS share of the supercomputer Hazel Hen at the High Performance Computing Center Stuttgart (HLRS), as well as on the machines of the Max Planck Computing and Data Facility (MPCDF) in Garching, Germany.

\\

We acknowledge the Virgo Consortium for making their simulation data available. The EAGLE simulations were performed using the DiRAC-2 facility at Durham, managed by the ICC, and the PRACE facility Curie based in France at TGCC, CEA, Bruyères-le-Châtel.
      
\end{acknowledgements}

\section*{Data availability}

All observational and simulation data are publicly available, except the SIMBA data generated by \citet{Akins2022}. The JWST/NIRCam data is available at the DAWN JWST Archive at \url{https://dawn-cph.github.io/dja/blog/2024/08/16/morphological-data/}. The photometric CANDELS/3D-HST catalogs (available at \url{https://archive.stsci.edu/prepds/3d-hst/}) are described by \citet[we use version 4.1]{Skelton2014}. The TNG100 subhalo and particle properties as well as the dust-attenuated fluxes created by \citet{Nelson2018} are available at \url{https://www.tng-project.org/} (described by \citealt{Nelson2019}). For EAGLE, the public database containing the dust-attenuated fluxes from \citet{Camps2018} is described by \citet{McAlpine2016} and available at \url{https://icc.dur.ac.uk/Eagle/database.php}. The geometrical models from DirtyGrid are described by \citet{Law2018}, available at \url{https://stsci.app.box.com/v/dirtygrid}. We make custom-made catalogs for all publicly available datasets as well as our own data products (broadband fluxes for TNG100 galaxies generated with SKIRT, EAZY-derived data for CANDELS/3D-HST galaxies with recent SED templates) and analysis scripts necessary to reproduce our results publicly available at \url{https://github.com/andreagebek/TNG100_UVJ}.

%
%

\bibliographystyle{aa}
\bibliography{main.bib}

\begin{appendix}

\section{Comparison to CANDELS/3D-HST}\label{sec:Observation variations}

\begin{figure*}
    \centering
    \includegraphics[width=\textwidth]{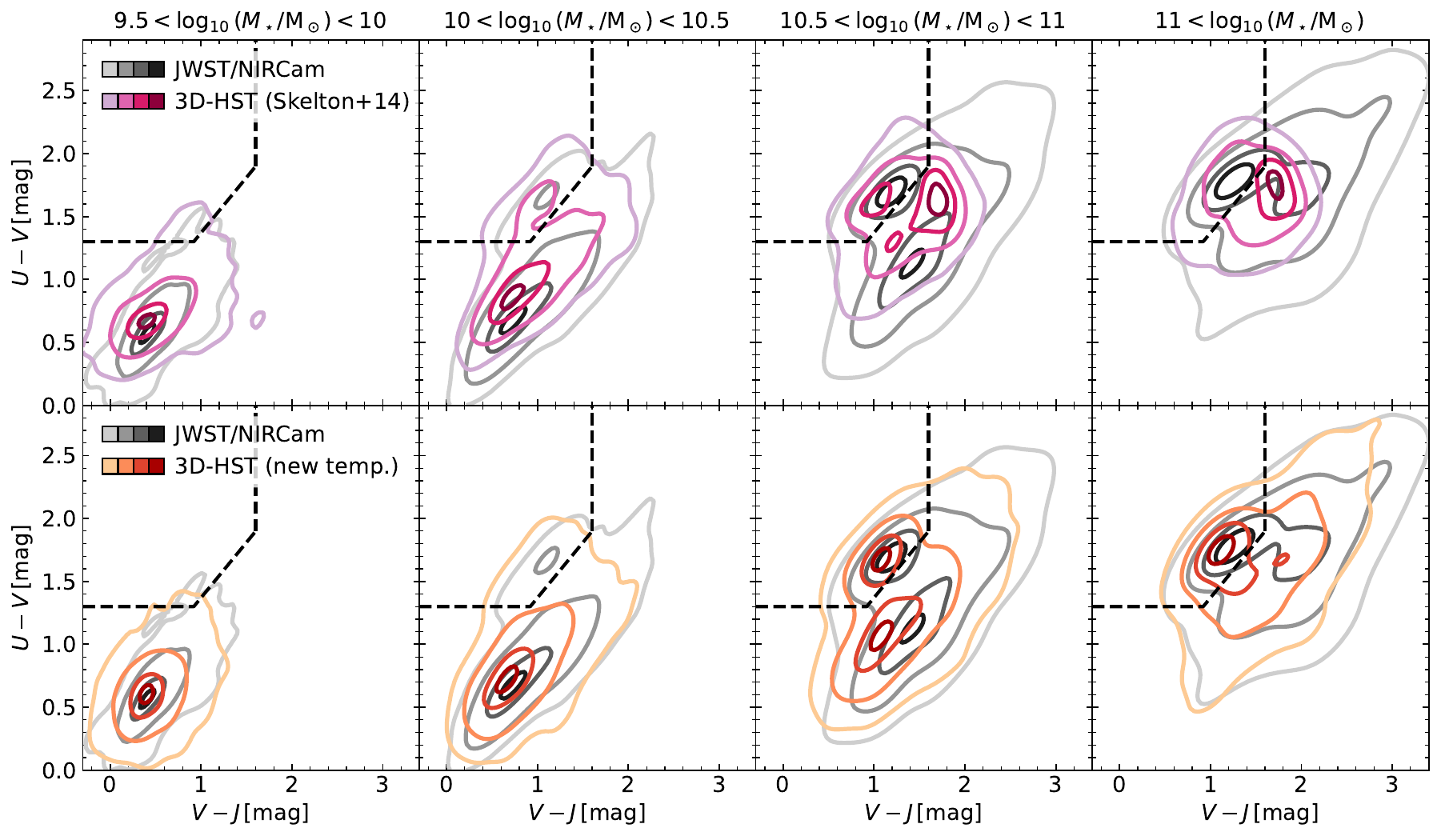}
    \caption{UVJ diagram for various observational data, for galaxies with $1.8\leq z\leq2.2$. Our fiducial observational dataset (JWST/NIRCam) is indicated by the grey contours. In the upper row, we show the UVJ colors from the CANDELS/3D-HST photometric catalogs from \citet{Skelton2014}. The orange contours in the lower row are based on exactly the same photometry as in \citet{Skelton2014}, but the rest-frame colors were rederived with a more recent set of EAZY templates. At low stellar masses, all observational datasets agree reasonably well. For more massive star-forming galaxies, however, the results from \citet{Skelton2014} deviate due to the lack of dusty SED templates in their photometric redshift-fitting procedure.}
    \label{fig:UVJ_obsDataComparison}
\end{figure*}

We have adopted a single observational dataset that incorporates recent JWST/NIRCam photometry throughout this study. Before JWST, the canonical observational dataset to study the UVJ diagram at cosmic noon for a statistical sample of galaxies was CANDELS/3D-HST (see e.g. \citealt{Fumagalli2014}; \citealt{Fang2018}; \citealt{Donnari2019}; \citealt{Zuckerman2021}; \citealt{Nagaraj2022}; \citealt{Akins2022}) due to the combination of spatial resolution with HST and broad wavelength coverage up to the rest-frame J-band with IRAC (onboard \textit{Spitzer}). In Fig.~\ref{fig:UVJ_obsDataComparison}, we compare the UVJ diagram from the publicly available\footnote{Version 4.1, available at \url{https://archive.stsci.edu/prepds/3d-hst/}} photometric CANDELS/3D-HST catalogs (\citealt{Skelton2014}) to our fiducial observational dataset. Both datasets are limited to $1.8<z<2.2$, additionally we apply the basic selection criteria (use\_phot flag of one, star\_flag of zero) suggested by \citet{Skelton2014} when using the CANDELS/3D-HST data.

The stellar masses, redshifts, and rest-frame colors from the CANDELS/3D-HST catalogs are derived in a similar way as our JWST/NIRCam data using EAZY. However, differences in the available bands, aperture photometry, and EAZY settings can give rise to systematic differences. At low stellar masses, we find that the observational datasets match reasonably well. At high stellar masses, however, the star-forming galaxies in the CANDELS/3D-HST data pile up at $V-J\approx1.8\,\mathrm{mag}$. This is a well-known effect arising due to the lack of dusty star-forming galaxy templates when EAZY was run on the CANDELS/3D-HST photometric data (see also Appendix C of \citealt{Whitaker2010}).

To test this, we  reran EAZY on the photometric catalogs from \citet{Skelton2014} using a more recent set of EAZY templates\footnote{We used the SED template parameter file `blue\_sfhz\_13.param', which can be found at \url{https://github.com/gbrammer/eazy-photoz/} (git commit hash 4747b59). Thirteen templates were generated with the Flexible Stellar Population Synthesis code (\citealt{Conroy2009}; \citealt{Conroy2010}), and a fourteenth empirical template is added to account for extreme emission lines (see Sect. 2.2.3 in \citealt{Weibel2024}).}. These results are shown by the red contours in Fig.~\ref{fig:UVJ_obsDataComparison}. We emphasize that the only difference between the CANDELS/3D-HST data in the upper and lower panels of Fig.~\ref{fig:UVJ_obsDataComparison} is due to the SED templates used in EAZY, as the underlying photometric data is exactly the same. With the newer templates, we find that the distribution of massive, star-forming galaxies in CANDELS/3D-HST now reaches much redder $\textit{V}-\textit{J}$ colors, as seen in the JWST/NIRCam data. Some smaller systematic differences between CANDELS/3D-HST and JWST/NIRCam persist (especially for $M_\star>10^{10.5}\,\mathrm{M}_\odot$), but we consider the agreement between the observational datasets to be satisfactory.

\section{SKIRT post-processing variations}\label{sec:SKIRT variations}

\begin{figure*}
    \centering
    \includegraphics[width=\textwidth]{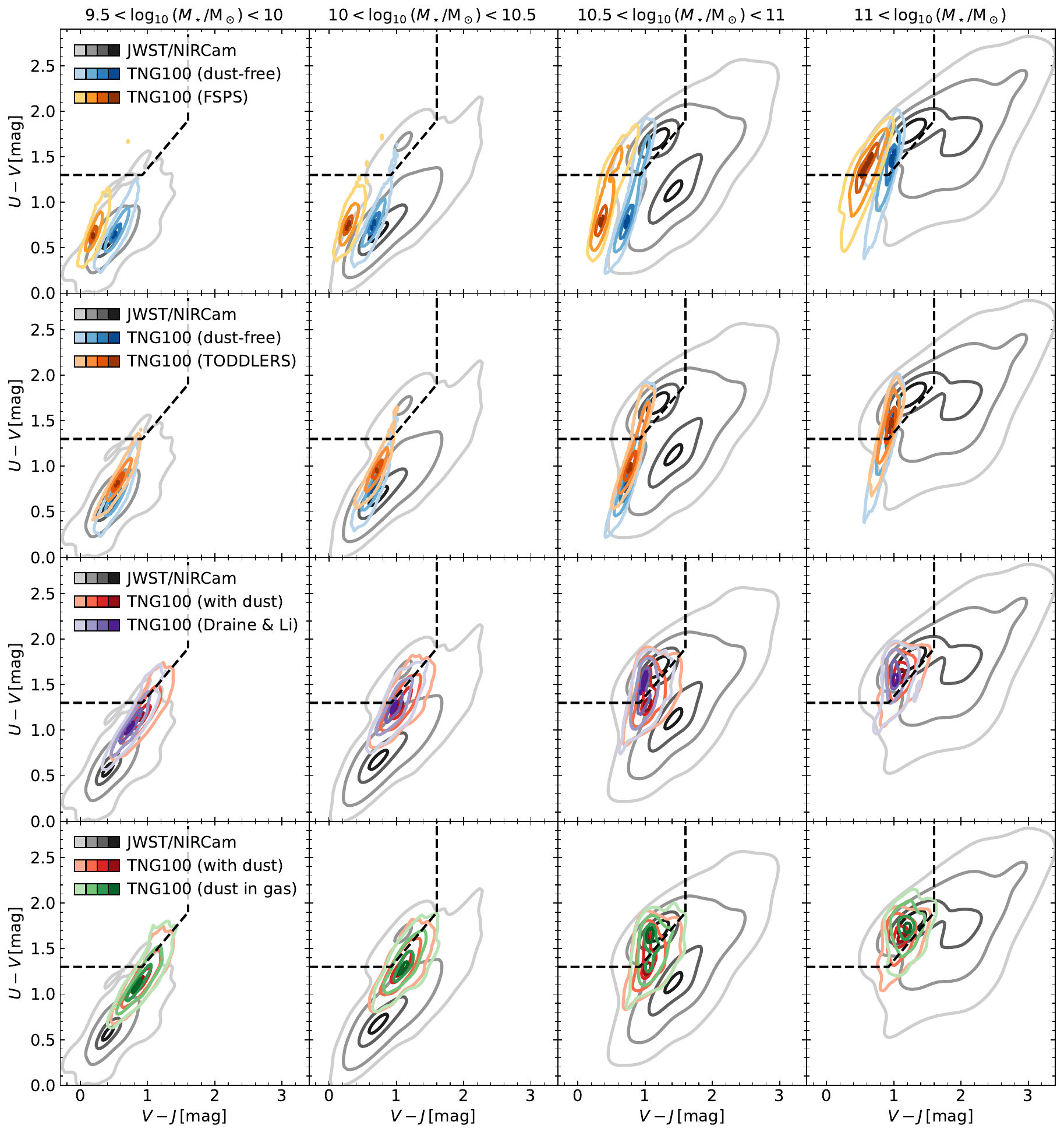}
    \caption{ Impact of variations in our SKIRT post-processing scheme for the TNG100 galaxies. In the upper two rows, the blue contours show our fiducial dust-free TNG100 colors, obtained with the BPASS templates. The orange contours in the first row indicate the UVJ colors obtained when using FSPS-MILES instead of BPASS. In the second row, we add local (unresolved) dust by using the TODDLERS templates for star-forming regions for all stellar populations with ages below 30\,Myr (older stellar populations are still modeled with BPASS). The lower two rows show variations in our dust modeling. Our fiducial post-processing scheme with a dust-to-metal ratio of 0.5, the THEMIS dust model, and assigning dust only to ISM gas cells according to the criterion of \citep{Torrey2012,Torrey2019} is indicated by the red contours. In the third row, we use the dust model from \citet{Draine2007} instead of THEMIS for the diffuse (resolved) dust in the ISM (purple contours). The fourth row shows the UVJ colors obtained when assigning dust to all gas cells that are gravitationally bound to the galaxy (green contours), as opposed to just the ISM gas cells. We find that, except for the choice of stellar population templates (first row), the impact of these SKIRT post-processing variations on the simulated TNG100 colors are generally minor (see main text for more information).}
    \label{fig:UVJ_SKIRTvariations}
\end{figure*}

Throughout this study, we have assumed a fiducial post-processing setup within SKIRT to compute broadband fluxes for TNG100 galaxies. This fiducial setup consists of the usage of the BPASS templates to generate dust-free fluxes. We compare them to dust-free fluxes obtained with the FSPS-MILES templates in the first row of Fig.~\ref{fig:UVJ_SKIRTvariations}. BPASS yields substantially redder $\textit{V}-\textit{J}$ colors than FSPS-MILES by $\lesssim0.37\,\mathrm{mag}$ (see footnote~\ref{footnote} on how we quote differences between various TNG100-SKIRT color distributions), while the $\textit{U}-\textit{V}$ colors are only marginally redder ($\lesssim0.06\,\mathrm{mag}$). We have also tested the SED template library from \citet{Bruzual2003}, and found that it yields almost exactly the same UVJ colors as FSPS-MILES.

For dust-attenuated fluxes, we used the TODDLERS templates for all stellar populations with ages below 30\,Myr to model unresolved dust in star-forming regions. We show the impact of replacing BPASS with TODDLERS for the young stellar populations in the second row of Fig.~\ref{fig:UVJ_SKIRTvariations}. While the $\textit{V}-\textit{J}$ colors are hardly affected by the dust in the star-forming regions, $\textit{U}-\textit{V}$ becomes redder by $\lesssim0.19\,\mathrm{mag}$.

In our fiducial setup for the dust-attenuated fluxes, we also added dust in the diffuse ISM using a fixed dust-to-metal ratio of 50\,\%. In the third row of Fig.~\ref{fig:UVJ_SKIRTvariations}, we show the impact of using the dust model of \citet{Draine2007} instead of our fiducial choice (THEMIS). We find minor differences in the $\textit{V}-\textit{J}$ colors ($\lesssim0.13\,\mathrm{mag}$) and even smaller differences in $\textit{U}-\textit{V}$. We have also tested the dust model of \citet{Zubko2004}, which effectively gives the same UVJ colors as our fiducial post-processing setup. The \citet{Draine2007} and \citet{Zubko2004} dust models rely on theoretical calculations, unlike THEMIS which is based on laboratory measurements. However, all three dust models are tuned to reproduce Milky Way observations. We find that the impact of dust model variations is marginal, but we caution that these results are based on dust models that are all anchored to Milky Way data. To what extent this finding can be extrapolated to dust properties at $z=2$ is unclear at present.

When assigning dust to gas cells for TNG100 galaxies, we use a temperature and density-dependent criterion (\citealp{Torrey2012,Torrey2019}) to determine the dust-containing ISM. The hotter, lower-density gas phase does not receive any dust in our fiducial method. In the lower panels of Fig.~\ref{fig:UVJ_SKIRTvariations}, we show the UVJ diagram for TNG100 galaxies when assigning dust to all gas cells (green contours), as opposed to the fiducial method where we assign dust only to ISM gas cells (red contours). Assigning dust to all gas cells leads to only marginally redder UVJ colors ($\lesssim0.08\,\mathrm{mag}$).

Overall, we conclude that our SKIRT-derived UVJ colors for TNG100 galaxies are robust against most variations in the post-processing setup. The only exception is the choice of SED templates, where we find that BPASS gives significantly redder $\textit{V}-\textit{J}$ colors\footnote{According to \citet{Stanway2018}, the different optical-NIR colors stem from varying treatments of the asymptotic giant branch phase and different stellar atmosphere models.} than other template libraries. We note that this means that the usage of any other template library would cause even more significant tension with observational data than we encountered (e.g., Fig.~\ref{fig:massiveSFgalaxies_hist}), reinforcing our results from Sect.~\ref{sec:DSFGs}.

\end{appendix}

\end{document}